\documentclass[prd,onecolumn,amsmath,amssymb,floatfix,nofootinbib]{revtex4}
\pdfoutput=1

\usepackage{amssymb}
\usepackage{amsmath}
\usepackage{feynmp}
\usepackage{slashed}

\usepackage{caption}
\usepackage{subfig}

\usepackage{braket}
\usepackage{graphicx}
\usepackage{graphics}
\usepackage{dcolumn}
\usepackage{color}
\usepackage{rotate}
\usepackage{fancyhdr}
\usepackage{indentfirst}
\usepackage{umoline}

\usepackage{hyperref}
\hypersetup{
    colorlinks=true,
    linkcolor=black,
    filecolor=black,
    urlcolor=black,
}

\def\be{\begin{eqnarray}}
\def\ee{\end{eqnarray}}
\def\no{\nonumber}

\newcommand{\abs}[1]{\lvert#1\rvert}

\def\lsim{\mathrel{\rlap{\lower4pt\hbox{\hskip1pt$\sim$}}
     \raise1pt\hbox{$<$}}}         
\def\gsim{\mathrel{\rlap{\lower4pt\hbox{\hskip1pt$\sim$}}
     \raise1pt\hbox{$>$}}}         
\newcommand{\beq}{\begin{equation}}
\newcommand{\eeq}{\end{equation}}

\newenvironment{Eqnarray}{\arraycolsep 0.14em\begin{eqnarray}}{\end{eqnarray}}
\def\beqa{\begin{Eqnarray}}
\def\eeqa{\end{Eqnarray}}

\begin{document}
\title{$\Upsilon$ and $\psi$ leptonic decays as probes of solutions to the $R\left(D^{(*)}\right)$ puzzle}

\author{Daniel Aloni$^{1a}$, Aielet Efrati$^{1a}$, Yuval Grossman$^{2b}$ and Yosef Nir$^{1a}$}
\affiliation{$^1$Department of Particle Physics and Astrophysics, Weizmann Institute of Science, Rehovot, Israel 7610001 \\
$^2$Department of Physics, LEPP, Cornell University, Ithaca, NY 14853}
\email{$^a$daniel.aloni, aielet.efrati, yosef.nir@weizmann.ac.il, $^b$yg73@cornell.edu}

\begin{abstract}
\noindent
Experimental measurements of the ratios $R(D^{(*)})\equiv\frac{\Gamma(B\to D^{(*)}\tau\nu)}{\Gamma(B\to D^{(*)}\ell\nu)}$ ($\ell=e,\mu$) show a $3.9\sigma$ deviation from the Standard Model prediction. In the absence of light right-handed neutrinos, a new physics contribution to $b\to c\tau\nu$ decays necessarily modifies also $b\bar b\to\tau^+\tau^-$ and/or $c\bar c\to\tau^+\tau^-$ transitions. These contributions lead to violation of lepton flavor universality in, respectively, $\Upsilon$ and $\psi$ leptonic decays. We analyze the constraints resulting from measurements of the leptonic vector-meson decays on solutions to the $R(D^{(*)})$ puzzle. Available data from BaBar and Belle can already disfavor some of the new physics explanations of this anomaly. Further discrimination can be made by measuring $\Upsilon(1S,2S,3S)\to\tau\tau$ in the upcoming Belle II experiment.
\end{abstract}

\maketitle

\section{Introduction}
The Standard Model (SM) predicts that the electroweak interactions respect lepton flavor universality (LFU). Violation of LFU, beyond the small effects of the Yukawa interactions (or, equivalently, of the charged lepton masses), will constitute clear evidence for physics beyond the SM. There is growing experimental evidence that LFU is broken in
the ratios
\beq
R(D^{(*)})\equiv\frac{\Gamma(B\to D^{(*)}\tau\nu)}{\Gamma(B\to D^{(*)}\ell\nu)},\ \ \ (\ell=e,\mu).
\eeq
The combined results of BaBar~\cite{Lees:2012xj,Lees:2013uzd}, Belle~\cite{Huschle:2015rga,Sato:2016svk,Abdesselam:2016xqt,Hirose:2016wfn} and LHCb~\cite{Aaij:2015yra} read~\cite{Amhis:2016xyh}
\beqa\label{eq:expR}
R(D)=0.403\pm 0.047,\;\;\;
R(D^*)=0.310\pm 0.017,\;\;\;\rho=-0.23,
\eeqa
where $\rho$ is the experimental correlation between $R(D)$ and $R(D^*)$. (The updated Belle result, $R(D^*)=0.270\pm0.044$~\cite{Hirose:2016wfn}, is not included in the HFAG average~\cite{Amhis:2016xyh}.)
The SM predictions are~\cite{Aoki:2016frl,Fajfer:2012vx}
\beqa
R(D)=0.300\pm0.008,\;\;\;
R(D^*)=0.252\pm0.003.
\eeqa
Combined, these results show a deviation from the SM prediction at a level of $3.9\sigma$~\cite{Amhis:2016xyh}.

This $R(D^{(*)})$ puzzle received a lot of attention in the literature in recent years. Various works analyze the deviation in terms of effective field theory (EFT), construct explicit models of new physics (NP), and introduce new observables and kinematical distributions which can shed light on the flavor and Lorentz structure of the underlying theory \cite{Sakaki:2012ft,Fajfer:2012jt,Crivellin:2012ye,Datta:2012qk,Choudhury:2012hn,Celis:2012dk,Tanaka:2012nw,Biancofiore:2013ki,Duraisamy:2013kcw,Dorsner:2013tla,Sakaki:2013bfa,Duraisamy:2014sna,Bhattacharya:2014wla,Alonso:2015sja,Greljo:2015mma,Calibbi:2015kma,Freytsis:2015qca,Ivanov:2015tru,Bhattacharya:2015ida,Bauer:2015knc,Hati:2015awg,Fajfer:2015ycq,Barbieri:2015yvd,Cline:2015lqp,Alonso:2016gym,Dorsner:2016wpm,Boucenna:2016wpr,Buttazzo:2016kid,Das:2016vkr,Nandi:2016wlp,Li:2016vvp,Feruglio:2016gvd,Alok:2016qyh,Ivanov:2016qtw,Becirevic:2016yqi,Sahoo:2016pet,Faroughy:2016osc,Bhattacharya:2016mcc,Ligeti:2016npd,Bardhan:2016uhr,Li:2016pdv,Barbieri:2016las,Alonso:2016oyd,Choudhury:2016ulr,Celis:2016azn,Ivanov:2017mrj,Dutta:2017xmj,Wei:2017ago,lambda_b2}.

Given the SM particle content and, in particular, assuming that the light neutrinos are purely left-handed, NP contributions to the $b\to c\tau\nu$ transition imply that NP contributions to $b\bar b\to\tau^+\tau^-$ and/or $c\bar c\to\tau^+\tau^-$ transitions are unavoidable. This point was discussed in Ref.~\cite{Faroughy:2016osc} which analyze the high $P_T$ distribution of the $\tau^+\tau^-$ signature at the LHC, arising from sea $b\bar b$ and/or $c\bar c$ annihilation. In this work we study new observables which are governed by the $c\bar c\to\tau\tau$ and $b\bar b\to\tau\tau$ transitions: non-universality in leptonic decays of $\psi$ and $\Upsilon$ quarkonia. Specifically, we study the ratios
\beqa
R^V_{\tau/\ell}\equiv\frac{\Gamma\left(V\to\tau^+\tau^-\right)}{\Gamma\left(V\to\ell^+\ell^-\right)},\ \ \ (V=\psi,\Upsilon;\ \ \ell=e,\mu),
\eeqa
making two assumptions:
\begin{itemize}
\item The deviation of $R(D^{(*)})$ from the SM prediction is generated by gauge invariant effective operators of dimension six.
\item $R(D^{(*)})$ is modified by new physics that affects only the $B\to D^{(*)}\tau\nu$ decays (and not the $B\to D^{(*)}\ell\nu$ decays). Correspondingly, $R^V_{\tau/\ell}$ is modified because $V\to\tau\tau$ is affected (and not $V\to\ell\ell$).
\end{itemize}
We compare our results to the current and future sensitivity for LFU violation in leptonic decays of $\Upsilon$ and $\psi$ vector-mesons.

One advantage of using the relation between operators responsible for the $R(D^{(*)})$ anomaly and those modifying the leptonic decays of $\Upsilon$ and $\psi$ is that all processes occur at the same energy scale. Therefore once we fix the Wilson coefficients at the low scale to give the measured best fit values of $R(D^{(*)})$,
no RGE effects and mixing among other operators affect our predictions for $R^V_{\tau/\ell}$. 
Once an anomaly is found in these leptonic vector meson decays, a full UV model should be scrutinized, including a proper UV matching and RGE mixing, as well as its compatibility with other relevant observables.

Tests of LFU have been carried out for $\Upsilon(1S)$, $\Upsilon(2S)$ and $\Upsilon(3S)$~\cite{Besson:2006gj,delAmoSanchez:2010bt}. We collect these results in Table~\ref{tab:exp}. For $\psi(2S)$ the leptonic branching fractions are measured to be~\cite{Olive:2016xmw}:
\beqa
BR(\psi(2S)\to\tau^+\tau^-)&=&(3.1\pm0.4)\times10^{-3},\no\\
BR(\psi(2S)\to\mu^+\mu^-)&=&(7.9\pm0.9)\times10^{-3},\no\\
BR(\psi(2S)\to e^+e^-)&=&(7.89\pm0.17)\times10^{-3}.
\eeqa
The corresponding ratio is presented in Table~\ref{tab:exp}. We do not consider $\psi(1S)$ whose mass is below the $\tau^+\tau^-$ threshold. We also do not consider $\psi(3770)$ and $\Upsilon(4S)$ which have negligible branching fractions into leptons because their masses are above the $D\bar{D}$ and $B\bar{B}$ threshold, respectively.
\begin{table}[h!]
\caption{ \it Experimental results and SM predictions for $R^V_{\tau/\ell}$.}
\label{tab:exp}
\begin{center}
\begin{tabular}{|c|c|c|} \hline\hline
\rule{0pt}{1.0em}%
$V(nS)$ & SM prediction & Exp. value $\pm\sigma_{\rm stat}\pm\sigma_{\rm syst}$ \\[2pt] \hline\hline
$\Upsilon(1S)$ & $0.9924\pm\mathcal{O}(10^{-5})$ & $1.005\pm0.013\pm0.022$  \rule{0pt}{1.0em}\\
$\Upsilon(2S)$ & $0.9940\pm\mathcal{O}(10^{-5})$ & $1.04\pm0.04\pm0.05$   \rule{0pt}{1.0em}\\
$\Upsilon(3S)$ & $0.9948\pm\mathcal{O}(10^{-5})$ & $1.05\pm0.08\pm0.05$  \rule{0pt}{1.0em}\\
$\psi(2S)$ & $0.390\pm\mathcal{O}(10^{-4})$ & $0.39\pm0.05$ \rule{0pt}{1.0em} \\
\hline\hline
\end{tabular}
\end{center}
\end{table}

Within the SM, the QED partial decay widths of a vector quarkonium into charged lepton pairs obey~\cite{VanRoyen:1967nq}
\beq\label{eq:R_QED}
R^V_{\tau/\ell}\simeq(1+2x^2_\tau)(1-4x_\tau^2)^{1/2}
\eeq
where $x_\tau=m_\tau/m_V$. This approximation neglects the electron and muon masses, one-loop corrections and weak-current effects. This phase space factor is the leading source of flavor non-universality in the SM. The dominant corrections to this factor are at the level of $0.006x_\tau^2$, and arise from QED vertex corrections. A full discussion on the SM decay rate is presented in App.~\ref{app:SM}. The relevant masses are known with a great accuracy~\cite{Olive:2016xmw}:
\beqa
m_{\Upsilon(1S)}&=&9.46030\pm0.00026\ {\rm GeV},\no\\
m_{\Upsilon(2S)}&=&10.02326\pm0.00031\ {\rm GeV},\no\\
m_{\Upsilon(3S)}&=&10.3552\pm0.0005\ {\rm GeV},\no\\
m_{\psi(2S)}&=&3.686097\pm0.000025\ {\rm GeV},\no\\
m_\tau&=&1.77686\pm0.00012\ {\rm GeV}.
\eeqa
The non-universality predicted in the SM agrees very well with the experimental results, as is evident from Table~\ref{tab:exp}.

LFU in Upsilon decays was discussed in the literature in the context of light pseudo-scalar (see Refs.~\cite{SanchisLozano:2006gx,Domingo:2009tb} and references within). In this scenario, the radiative $\Upsilon\to\gamma\eta_b$ decay is followed by a mixing between the $\eta_b$ state and a CP-odd scalar $A$, for which the leptonic couplings are non-universal. Lepton flavor changing decays of heavy vector-mesons were discussed in
Ref.~\cite{Bhattacharya:2016mcc,Hazard:2016fnc}. While such decays are not directly relevant to our study, the formalism is similar.

The plan of this paper is as follows. The formalism for $V\to\ell\ell$ decays is introduced in Section \ref{sec:vll}. The effective field theory that is relevant to $R(D^{(*)})$ and to $R^V_{\tau/\ell}$ is introduced in Section \ref{sec:EFT}. In Section \ref{sec:results} we analyze a series of simplified models, where we add to the SM a single new boson. For each model, we find the numerical range for the Wilson coefficients that explain $R(D^{(*)})$, and obtain the resulting predictions for $R^V_{\tau/\ell}$. In Section \ref{sec:prospects} we compare these predictions to present measurements and discuss the prospects for improving the experimental accuracy in the future. Our conclusions are summarized in Section \ref{sec:concs}.

\section{$V\to\ell\ell$ decay rate}
\label{sec:vll}
The most general $V\to\ell^+\ell^-$ decay amplitude can be written as
\beqa
\mathcal{M}(V\to\ell^+\ell^-)=\left(\frac{f_V}{m_V}\right)\bar u(p_1,s_1)\left[A_V^{q\ell}\gamma_\mu+B^{q\ell}_V\gamma_\mu\gamma_5+\frac{C_V^{q\ell}}{m_V}\left(p_2-p_1\right)_\mu+\frac{iD_V^{q\ell}}{m_V}\left(p_2-p_1\right)_\mu\gamma_5\right]v(p_2,s_2)\epsilon^\mu(p),
\eeqa
where $A_V^{q\ell},\,B_V^{q\ell},\,C_V^{q\ell},\,D_V^{q\ell}$ are dimensionless parameters which depend on the Wilson coefficients of the operators controlling the $V\to\ell^+\ell^-$ decays at the perturbative level, and on meson-to-vacuum matrix elements at the non-perturbative level. The form-factors of the Lorentz structures $(p_2+p_1)_\mu$ and $(p_2+p_1)_\mu\gamma_5$ do not contribute to the rate. The decay width and $R^V_{\tau/\ell}$ are given by
\beqa
\Gamma[V\to\ell\ell]=&&\frac{f_V^2}{4\pi m_V}\sqrt{1-4x_\ell^2}\left[\abs{A_V^{q\ell}}^2\left(1+2x_\ell^2\right)+\abs{B_V^{q\ell}}^2\left(1-4x_\ell^2\right)+\frac{\abs{C_V^{q\ell}}^2}{2}\left(1-4x_\ell^2\right)^2+\frac{\abs{D_V^{q\ell}}^2}{2}\left(1-4x_\ell^2\right)\right.\no\\
&&\left.+2{\rm Re}\left[A_V^{q\ell} C_V^{*q\ell}\right]x_\ell\left(1-4x_\ell^2\right)\right]
,\no\\
R^V_{\tau/\ell}\simeq&&\frac{\sqrt{1-4x_\tau^2}}{\abs{A_V^{\ell,{\rm SM}}}^2}\left[\abs{A_V^{q\tau}}^2\left(1+2x_\tau^2\right)+\abs{B_V^{q\tau}}^2\left(1-4x_\tau^2\right)+\frac{\abs{C_V^{q\tau}}^2}{2}\left(1-4x_\tau^2\right)^2+\frac{\abs{D_V^{q\tau}}^2}{2}\left(1-4x_\tau^2\right)\right.\no\\
&&\left.+2{\rm Re}\left[A_V^{q\tau} C_V^{*q\tau}\right]x_\tau\left(1-4x_\tau^2\right)\right].
\eeqa
Within the SM, $A_V^{\ell,{\rm SM}}\simeq-4\pi\alpha Q_q$ and $B_V^{\ell,{\rm SM}},\,C_V^{\ell,{\rm SM}},\,D_V^{\ell,{\rm SM}}\simeq0$.
For the SM calculations we include QED one-loop correction which is further discussed in App.~\ref{app:SM}.

In order to calculate the $V\to\ell^+\ell^-$ decay rate in a specific UV or EFT model, one needs to find the relation between $A_V,B_V,C_V$ and $D_V$ and the Lagrangian parameters.
Including only terms which are relevant for a leptonic meson decay, we consider the following effective Lagrangian:
\beqa
\mathcal{L}_{\ell q}=&&
C_{VRR}^{q\ell  }\bar e_R\gamma^\mu e_R\bar q_R\gamma_\mu q_R
+C_{VRL}^{q\ell  }\bar e_R\gamma^\mu e_R\bar q_L\gamma_\mu q_L
+C_{VLR}^{q\ell  }\bar e_L\gamma^\mu e_L\bar q_R\gamma_\mu q_R
+C_{VLL}^{q\ell  }\bar e_L\gamma^\mu e_L\bar q_L\gamma_\mu q_L\no\\
+&&\left[C_{T}^{q\ell  }\bar e_L\sigma^{\mu\nu} e_R\bar q\sigma_{\mu\nu}q+C_{SL}^{q\ell}\bar e_Re_L\bar q_R q_L +C_{SR}^{q\ell}\bar e_Re_L\bar q_Lq_R+{\rm h.c.}\right]~.
\eeqa
We find:
\beqa
A_V^{q\ell}&&=-4\pi\alpha Q_q+\frac{m_V^2}{4}\left[\left(C_{VLL}^{q\ell}+C_{VRR}^{q\ell}+C_{VLR}^{q\ell}+C_{VRL}^{q\ell}\right)
+16x_\ell\frac{f_V^T}{f_V}{\rm Re}\left[C_{T}^{q\ell}\right]
\right],\no\\
B_V^{q\ell}&&=\frac{m_V^2}{4}\left(C_{VRR}^{q\ell}+C_{VRL}^{q\ell}-C_{VLR}^{q\ell}-C_{VLL}^{q\ell}\right),\no\\
C_V^{q\ell}&&=2m_V^2\frac{f_V^T}{f_V}{\rm Re}\left[C_{T}^{q\ell}\right],\no\\
D_V^{q\ell}&&=2m_V^2\frac{f_V^T}{f_V}{\rm Im}\left[C_{T}^{q\ell}
\right].
\eeqa
We introduced the form factors, $f_V$ and $f_V^T$, that are defined via the standard parametrization:
\beqa\label{eq:HadronicFF}
\bra{0}\bar q\gamma^\mu q\ket{V(p)}&&=f_Vm_V\epsilon^\mu(p),\no\\
\bra{0}\bar q\sigma^{\mu\nu}q\ket{V(p)}&&=if_V^T\left[\epsilon^\mu(p)p^\nu-\epsilon^\nu(p)p^\mu\right],
\eeqa
with $\sigma_{\mu\nu}=i[\gamma_\mu,\gamma_\nu]/2$, and
$\bra{0}\bar qq\ket{V(p)}=\bra{0}\bar q\gamma_5q\ket{V(p)}=\bra{0}\bar q\gamma^\mu\gamma_5 q\ket{V(p)}=0$. The relevant ratio $f_V^T/f_V$ should be determined from measurements or lattice calculations. In the heavy quark limit $f_V=f_V^T$. This is an excellent approximation for the $\Upsilon$ meson. For the $\psi(2S)$ state, however, relativistic effects correct this relation by a few percent~\cite{Becirevic:2013bsa}. We checked that this is a sub-leading effect, therefore in the following we neglect this correction.

\section{The Effective Field Theory}\label{sec:EFT}
Assuming that the NP contributions are related to heavy degrees of freedom, their effects can be presented by non-renormalizable terms in the Lagrangian. There are eight combinations of two lepton and two quark fields that can be contracted into $SU(3)_C\times SU(2)_L\times U(1)_Y$ and Lorentz invariant operators:
\beqa
\bar L L\bar Q Q,\;\;\;\bar eL\bar uQ,\;\;\;\bar eL\bar Qd,\;\;\;\bar LL\bar uu,\;\;\;\bar LL\bar dd,\;\;\;\bar ee\bar uu,\;\;\;\bar ee\bar dd,\;\;\;\bar ee\bar QQ,
\eeqa
where $L$ and $e$ are the $SU(2)_L$ doublet and singlet lepton fields, and $Q$, $u$ and $d$ are the $SU(2)_L$ doublet, up-singlet and down-singlet quark fields. Only the first three combinations can introduce the charged-current interaction needed for a $b\to c$ transition. Specifying the Lorentz and $SU(2)_L$ contractions, we write the complete set of (linearly dependent) gauge-invariant operators in Table~\ref{tab:operators}. Wherever possible, we follow the notations of Ref.~\cite{Freytsis:2015qca}.

Given that the experimental central values of $R(D^{(*)})$ deviate by order $30\%$ from the SM predictions, and given that the SM amplitude is tree-level and only mildly CKM-suppressed, it is likely that the NP contribution which accounts for the deviation is also tree-level. It is instructive, therefore, to consider simplified UV models, each with a single new (scalar or vector) boson. Following Ref.~\cite{Freytsis:2015qca}, we specify the NP field content which generates the operators in Table~\ref{tab:operators}. Our convention is such that $\psi^c=-i\gamma^2\psi^*$.
\begin{table}[h!]
\caption{ \it Dimension six four-fermion Operators}
\label{tab:operators}
\begin{center}
\begin{tabular}{|c|c|c|c|} \hline\hline
\rule{0pt}{1.0em}%
Field content 	& Operator &  Fierz identities & NP rep, $s=0,1$ \\[2pt] \hline\hline
\rule{0pt}{1.2em}%
$\bar L L\bar Q Q$ 	
& $\left(\bar L\gamma^\mu L\right)\left(\bar Q\gamma_\mu Q\right)$ &   $\mathcal{O}_{V_L}^1$ &  $B^\prime_\mu\sim(1,1)_0$  \rule{0pt}{1.0em}\\
& $\left(\bar L\gamma^\mu\tau_a L\right)\left(\bar Q\gamma_\mu\tau^a Q\right)$ & $\mathcal{O}_{V_L}^2$ & $W^\prime_\mu\sim(1,3)_0$  \rule{0pt}{1.0em}\\
& $\left(\bar L\gamma^\mu Q\right)\left(\bar Q\gamma_\mu L\right)$  & $\frac{1}{2}\mathcal{O}_{V_L}^1+2\mathcal{O}_{V_L}^2$ &  $U_\mu\sim(3,1)_{+2/3}$ \rule{0pt}{1.0em}\\
& $\left(\bar L\gamma^\mu \tau_aQ\right)\left(\bar Q\gamma_\mu \tau^aL\right)$ & $\frac{3}{8}\mathcal{O}_{V_L}^1-\frac{1}{2}\mathcal{O}_{V_L}^2$ & $X_\mu\sim(3,3)_{+2/3}$ \rule{0pt}{1.0em}\\
& $\left(\bar L \epsilon^T Q^c\right)\left(\overline{Q^c}\epsilon L\right)$ & $\frac{1}{4}\mathcal{O}_{V_L}^1 - \mathcal{O}_{V_L}^2$ &  $S\sim(3,1)_{-1/3}$ \rule{0pt}{1.0em} \\
&  $\left(\bar L  \tau_a\epsilon^T Q^c\right)\left(\overline{Q^c}\epsilon\tau^a L\right)$ & $\frac{3}{16}\mathcal{O}_{V_L}^1+\frac{1}{4}\mathcal{O}_{V_L}^2$ & $T\sim(3,3)_{-1/3}$ \rule{0pt}{1.0em}\\
& $\left(\bar L\epsilon^T\sigma^{\mu\nu}Q^c\right)\left(\overline{Q^c}\epsilon\sigma_{\mu\nu} L\right)$      & 0 & -- \rule{0pt}{1.0em}\\
& $\left(\bar L \tau_a\epsilon^T\sigma^{\mu\nu}Q^c\right)\left(\overline{Q^c}\epsilon\tau^a\sigma_{\mu\nu}L\right)$ & 0 & --
						\rule{0pt}{1.0em}\\ \hline\rule{0pt}{1.2em}
$ \bar eL\bar uQ$				
&  $\left(\bar eL\right)\epsilon\left(\bar uQ\right)$   & $\mathcal{O}_{S_L}$ &  $\phi\sim(1,2)_{+1/2}$ \rule{0pt}{1.0em}\\
&   $\left(\bar uL\right)\epsilon\left(\bar eQ\right)$   & $-\frac{1}{2}\mathcal{O}_{S_L}-\frac{1}{8}\mathcal{O}_{T}$ & $D\sim(3,2)_{+7/6}$ \rule{0pt}{1.0em}\\
&   $\left(\overline{L^c}\epsilon Q\right)\left(\bar eu^c\right)$   & $-\frac{1}{2}\mathcal{O}_{S_L}+\frac{1}{8}\mathcal{O}_{T}$ & $S\sim(3,1)_{-1/3}$  \rule{0pt}{1.0em} \\
&  $\left(\bar e\sigma^{\mu\nu}L\right)\epsilon \left(\bar u\sigma_{\mu\nu}Q\right)$  & $\mathcal{O}_{T}$ & --\rule{0pt}{1.0em} \\
&   $\left(\bar u\sigma^{\mu\nu}L\right)\epsilon \left(\bar e\sigma_{\mu\nu}Q\right)$  &  $-6\mathcal{O}_{S_L}+\frac{1}{2}\mathcal{O}_{T}$ & --\rule{0pt}{1.0em} \\
&   $\left(\overline{L^c}\sigma_{\mu\nu}\epsilon Q\right)\left(\bar e\sigma^{\mu\nu}u^c\right)$   & $6\mathcal{O}_{S_L}+\frac{1}{2}\mathcal{O}_{T}$ & --
				\rule{0pt}{1.0em}\\ \hline\rule{0pt}{1.2em}
$\bar eL\bar Q d$  &   $\left(\bar Qd\right)\left(\bar eL\right)$  & $\mathcal{O}_{S_R}$ & $\phi\sim(1,2)_{+1/2}$ \rule{0pt}{1.0em} \\
						&   $\left(\bar Q\gamma^\mu L\right)\left(\bar e\gamma_\mu d\right)$  & $-2\mathcal{O}_{S_R}$ & $U_\mu\sim(3,1)_{+2/3}$  \rule{0pt}{1.0em} \\
						&   $\left(\bar Q\gamma^\mu e^c\right)\left(\overline{d^c}\gamma_\mu L\right)$  & $2\mathcal{O}_{S_R}$ & $V_\mu\sim(3,2)_{-5/6}$  \rule{0pt}{1.0em} \\
						&   $\left(\bar Q\sigma^{\mu\nu}d\right)\left(\bar e\sigma_{\mu\nu}L\right)$   & 0 &  -- \rule{0pt}{1.0em}\\
\hline
$\bar LL\bar uu$
& $\left(\bar L\gamma^\mu L\right)\left(\bar u\gamma_\mu u\right)$ & $\mathcal{O}_{V_L}^3$ & $B^\prime_\mu\sim(1,1)_0$ \rule{0pt}{1.0em} \\
& $\left(\bar L\gamma^\mu u^c\right)\left(\overline{L^c}\gamma_\mu u\right)$ & $\mathcal{O}_{V_L}^3$ & $Y_\mu\sim(3,2)_{+1/6}$\rule{0pt}{1.0em} \\
& $\left(\bar Lu\right)\left(\bar uL\right)$ & $-\frac{1}{2}\mathcal{O}_{V_L}^3$ & $D\sim(3,2)_{+7/6}$\rule{0pt}{1.0em} \\
& $\left(\bar L\sigma^{\mu\nu}u\right)\left(\bar u\sigma_{\mu\nu}L\right)$ & 0 & -- \rule{0pt}{1.0em} \\
\hline\rule{0pt}{1.2em}
$\bar LL\bar dd$
& $\left(\bar L\gamma^\mu L\right)\left(\bar d\gamma_\mu d\right)$ &$\mathcal{O}_{V_L}^4$ & $B^\prime_\mu\sim(1,1)_0$  \rule{0pt}{1.0em} \\
& $\left(\bar L\gamma^\mu d^c\right)\left(\overline{L^c}\gamma_\mu d\right)$ & $\mathcal{O}_{V_L}^4$ & $V_\mu\sim(3,2)_{-5/6}$ \rule{0pt}{1.0em} \\
& $\left(\bar Ld\right)\left(\bar dL\right)$ & $-\frac{1}{2}\mathcal{O}_{V_L}^4$ & $D^\prime\sim(3,2)_{+1/6}$  \rule{0pt}{1.0em} \\
& $\left(\bar L\sigma^{\mu\nu}d\right)\left(\bar d\sigma_{\mu\nu}L\right)$ & 0 & -- \rule{0pt}{1.0em} \\
\hline\rule{0pt}{1.2em}
$\bar ee\bar QQ$
& $\left(\bar e\gamma^\mu e\right)\left(\bar Q\gamma_\mu Q\right)$ &$\mathcal{O}_{V_R}^1$ & $B^\prime_\mu\sim(1,1)_0$  \rule{0pt}{1.0em} \\
& $\left(\bar Q\gamma^\mu e^c\right)\left(\overline{Q^c}\gamma_\mu e\right)$ & $\mathcal{O}_{V_R}^1$ & $V_\mu\sim(3,2)_{-5/6}$ \rule{0pt}{1.0em} \\
& $\left(\bar Qe\right)\left(\bar eQ\right)$ & $-\frac{1}{2}\mathcal{O}_{V_R}^1$ & $D\sim(3,2)_{+7/6}$  \rule{0pt}{1.0em} \\
& $\left(\bar Q\sigma^{\mu\nu}e\right)\left(\bar e\sigma_{\mu\nu}Q\right)$ & 0 & -- \rule{0pt}{1.0em} \\
\hline\rule{0pt}{1.2em}
$\bar ee\bar uu$
& $\left(\bar e\gamma^\mu e\right)\left(\bar u\gamma_\mu u\right)$ &$\mathcal{O}_{V_R}^2$ & $B^\prime_\mu\sim(1,1)_0$  \rule{0pt}{1.0em} \\
& $\left(\bar e\gamma^\mu u\right)\left(\bar u\gamma_\mu e\right)$ & $\mathcal{O}_{V_R}^2$ & $Z_\mu\sim(3,2)_{+5/3}$ \rule{0pt}{1.0em} \\
& $\left(\bar eu^c\right)\left(\overline{u^c}e\right)$ & $\frac{1}{2}\mathcal{O}_{V_R}^2$ & $S\sim(3,1)_{-1/3}$  \rule{0pt}{1.0em} \\
& $\left(\bar e\sigma^{\mu\nu}u^c\right)\left(\overline{u^c}\sigma^{\mu\nu}e\right)$ & $0$ & --  \rule{0pt}{1.0em} \\
\hline\rule{0pt}{1.2em}
$\bar ee\bar dd$
& $\left(\bar e\gamma^\mu e\right)\left(\bar d\gamma_\mu d\right)$ &$\mathcal{O}_{V_R}^3$ & $B^\prime_\mu\sim(1,1)_0$  \rule{0pt}{1.0em} \\
& $\left(\bar e\gamma^\mu d\right)\left(\bar d\gamma_\mu e\right)$ &$\mathcal{O}_{V_R}^3$ & $U_\mu\sim(3,1)_{+2/3}$  \rule{0pt}{1.0em} \\
& $\left(\bar ed^c\right)\left(\overline{d^c}e\right)$ & $\frac{1}{2}\mathcal{O}_{V_R}^3$ & $S^\prime\sim(3,1)_{-4/3}$  \rule{0pt}{1.0em} \\
& $\left(\bar e\sigma^{\mu\nu}d^c\right)\left(\overline{d^c}\sigma^{\mu\nu}e\right)$ & $0$ & --  \rule{0pt}{1.0em} \\\hline\hline
\end{tabular}
\end{center}
\end{table}

As a linearly independent set, we choose the following four-fermion operators (written in the quark interaction basis):
\beqa\label{eq: linearly independent operator}
\mathcal{O}_{V_L}^1=&&
\left(\bar\nu_L\gamma^\mu \nu_L\right)\left(\bar u_L\gamma_\mu u_L\right)
+\left(\bar\nu_L\gamma^\mu \nu_L\right)\left(\bar d_L\gamma_\mu d_L\right)
+\left(\bar e_L\gamma^\mu e_L\right)\left(\bar u_L\gamma_\mu u_L\right)
+\left(\bar e_L\gamma^\mu e_L\right)\left(\bar d_L\gamma_\mu d_L\right),\no\\
\mathcal{O}_{V_L}^2=&&
\frac{1}{4}\left[\left(\bar\nu_L\gamma^\mu \nu_L\right)\left(\bar u_L\gamma_\mu u_L\right)
-\left(\bar\nu_L\gamma^\mu \nu_L\right)\left(\bar d_L\gamma_\mu d_L\right)
-\left(\bar e_L\gamma^\mu e_L\right)\left(\bar u_L\gamma_\mu u_L\right)
+\left(\bar e_L\gamma^\mu e_L\right)\left(\bar d_L\gamma_\mu d_L\right)\right]\no\\
&&+\frac{1}{2}\left[\left(\bar\nu_L\gamma^\mu e_L\right)\left(\bar d_L\gamma_\mu u_L\right)
+\left(\bar e_L\gamma^\mu\nu_L\right)\left(\bar u_L\gamma_\mu d_L\right)\right],\no\\
\mathcal{O}_{V_L}^3=&&
\left(\bar\nu_L\gamma^\mu \nu_L\right)\left(\bar u_R\gamma_\mu u_R\right)
+\left(\bar e_L\gamma^\mu e_L\right)\left(\bar u_R\gamma_\mu u_R\right),\no\\
\mathcal{O}_{V_L}^4=&&
\left(\bar\nu_L\gamma^\mu \nu_L\right)\left(\bar d_R\gamma_\mu d_R\right)
+\left(\bar e_L\gamma^\mu e_L\right)\left(\bar d_R\gamma_\mu d_R\right),\no\\
\mathcal{O}_{V_R}^1=&&
\left(\bar e_R\gamma^\mu e_R\right)\left(\bar u_L\gamma_\mu u_L\right)+\left(\bar e_R\gamma^\mu e_R\right)\left(\bar d_L\gamma_\mu d_L\right),\no\\
\mathcal{O}_{V_R}^2=&&
\left(\bar e_R\gamma^\mu e_R\right)\left(\bar u_R\gamma_\mu u_R\right),\no\\
\mathcal{O}_{V_R}^3=&&
\left(\bar e_R\gamma^\mu e_R\right)\left(\bar d_R\gamma_\mu d_R\right),\no\\
\mathcal{O}_{S_L}=&&
-\left(\bar e_R e_L\right)\left(\bar u_R u_L\right)
+\left(\bar e_R \nu_L\right)\left(\bar u_R d_L\right),\no\\
\mathcal{O}_{S_R}=&&
\left(\bar e_R \nu_L\right)\left(\bar u_L d_R\right)
+\left(\bar e_R e_L\right)\left(\bar d_L d_R\right),\no\\
\mathcal{O}_T=&&
\left(\bar e_R\sigma^{\mu\nu}\nu_L\right)\left(\bar u_R\sigma_{\mu\nu}d_L\right)
-\left(\bar e_R\sigma^{\mu\nu}e_L\right)\left(\bar u_R\sigma_{\mu\nu}u_L\right).
\eeqa

A comment is in order regarding other dimension six operators which may be generated by integrating out heavy particles. Two sets of operators relevant to the study of $R^V_{\tau/\ell}$ are operators of the form $\mathcal{O}_{Hl_{L,R}}=i\left(H^\dagger\overleftrightarrow{D}_\mu H\right)\left(\bar{l}_{L,R}\gamma^\mu l_{L,R}\right)$ and the dipole operators $\mathcal{O}_D= H\bar{L}\sigma^{\mu\nu}e_RF_{\mu\nu}$.

The $\mathcal{O}_{Hl}$ operators generate non-universal $Z$-mediated quarkonia decays, and further modify the $Z\to\tau\tau$ decay width. These two effects are related: The leading effect in $R_{\tau/\ell}^V$ arising from this operator is given by
\beqa
R^V_{\tau/\ell}\simeq R^{V,{\rm SM}}_{\tau/\ell}\left(1-\frac{m_V^2}{v^2}\frac{g_V^q}{4\pi\alpha Q_q}\left[\delta g_L^{Z\tau}+\delta g_R^{Z\tau}\right]\right),
\eeqa
where $v = 246$~GeV is the electroweak vacuum expectation value. The LEP measurements of $Z$-pole observables~\cite{ALEPH:2005ab} constrain this effect to be smaller than $10^{-5}$. For more details see App.~\ref{App:Zpole}.
These operators can be formed at the UV matching scale, in which case they are typically controlled by additional free parameters. They can be further generated by the mixing with four fermion operators, in which case fine-tuning cancelation may be needed to ensure that the $Z$-pole constraints are not violated.

Dipole operators might be generated at the UV by one loop processes or by RGE mixing with the four-fermi operator $\mathcal{O}_T$. Typically, the resulting contribution to the vector meson decay rate is $\mathcal{O}(\alpha v/m_V)$ compared to the contribution arising from the tensor operator. Yet, the Wilson coefficients generated in the mixing are $y_cV_{cb}$ suppressed~\cite{Alonso:2013hga,Pruna:2014asa}. Numerically, we find that this effect on $R_{\tau/\ell}^{\Upsilon(1S)}$ is always below the per-mil level and can be neglected. The dipole operators further modify the electric and magnetic dipole moments of the $\tau$ lepton.
For completeness we write the leading contributions of the dipole operators to $R^V_{\tau/\ell}$ and the relation to the taonic dipole moments in App.~\ref{App:Dpole}.

In general RGE effects generate also four-quark and four-lepton operators. Such effects were studied in {\it e.g.} Ref.~\cite{Feruglio:2016gvd} and we do not study their possible indirect implications on our analysis.

\section{Numerical results}\label{sec:results}

As explained above, it is useful to explore simplified UV models which generate the required four-fermion operators at tree-level. In the following subsections we examine various sets of effective operators formed by the integration out of heavy scalar and vector bosons which have the right quantum numbers to modify the $b\to c\tau\nu$ transition. For each case, we present the relevant UV couplings and obtain the resulting CC and NC operators (neglecting, as explained above, RGE effects). We find within $95\%$ C.L. the numerical values for the Wilson coefficients which minimize the observed anomaly in $R(D^{(*)})$, and the corresponding predictions for $R^V_{\tau/\ell}$ for the $\Upsilon$ and $\psi(2S)$ states.
%
Some of the simplified models we study cannot ease the tension between the theory and the $R(D^{(*)})$ measurements. Clearly, in these models the best fit point of the new couplings is zero, giving back the SM. Nevertheless, for completeness we also consider these models and present our results for them.
We present our main results in Table~\ref{tab:results}, where only models which ease the tension below $\chi^2 =9$ (corresponding to $3\sigma$ with one degree of freedom) are included.
Note that in some cases only one of the decay modes, {\it i.e.} $\Upsilon\to\tau\tau$ or $\psi\to\tau\tau$, is modified, due to the specific $SU(2)$ structure of the effective operators. Furthermore, in the 2HDM case $\left(\phi\sim(1,2)_{+1/2}\right)$ neither of these decays is modified because of the vector structure of the $q\bar q$ mesons.
\begin{table}[h!]
\caption{ \it The simplified (single boson) models and the predicted range for $R^V_{\tau/\ell}$ for $V=\Upsilon(1S),\psi(2S)$. The achievable and projected uncertainties are our estimations, see the text for more details.}
\label{tab:results}
\begin{center}
\begin{tabular}{|l|l|l|l|} \hline\hline
\rule{0pt}{1.2em} %
UV field content 	 &  $R^{\Upsilon(1S)}_{\tau/\ell}$ & $R^{\psi(2S)}_{\tau/\ell}$ & Predicted modification to $R^{\Upsilon(1S)}_{\tau/\ell}$ \\[2pt] \hline\hline
\rule{0pt}{1.2em}%
$W^\prime_\mu\sim(1,3)_0$
& 0.989-0.991 & 0.390 \rule{0pt}{1.0em}
& Decrease by $0.2\%-0.4\%$\\ \rule{0pt}{1.2em}
$U_\mu\sim(3,1)_{+2/3}$
&  0.952-0.990 & SM \rule{0pt}{1.0em}
& Decrease by $0.3\%-4.0\%$ \\ \rule{0pt}{1.2em}
%
%
$S\sim(3,1)_{-1/3}$
& SM & 0.389-0.390 \rule{0pt}{1.0em}
& -- \\ \rule{0pt}{1.2em}
%
%
%
%
$V_\mu\sim(3,2)_{-5/6}$
& 0.976-0.987 & SM  \rule{0pt}{1.0em}
& Decrease by $0.5\%-1.6\%$ \\ \hline\hline \rule{0pt}{1.2em}
SM  & $0.992$ & $0.390$ & \\ \rule{0pt}{1.2em}
Current measurement & $1.005\pm0.025$ & $0.39\pm0.05$ & \\ \rule{0pt}{1.2em}
Achievable uncertainty (with current data) & $\pm0.01$ & $\pm0.02$ & \\ \rule{0pt}{1.2em}
Projected uncertainty ($\mathcal{L}^{\Upsilon(3S)}=1/{\rm ab}$ in Belle II)  & $\pm0.004$ & -- & \\
 \hline \hline
\end{tabular}
\end{center}
\end{table}

One clear advantage of analyzing such simplified models is that each
scenario predicts distinct relations between the various EFT
operators. These relations can, however, be modified due to $SU(2)_L$
breaking effects. Let us explain why we ignore these effects in our analysis. Electroweak breaking effects split the spectrum of the charged and neutral NP fields, which, in turn, changes the relations between the Wilson coefficients of the CC and NC effective operators by $\mathcal{O}\left(\Delta M/M\right)$. This modifies our predictions for $R^V_{\tau/\ell}$. The unavoidable loop-induced splitting is smaller than a GeV, and therefore negligible. Tree-level splitting is typically of order $\Delta M/M\simeq v^2/M^2$ and might change our final results by a few percent. This splitting is, however, generated by a free parameter in the scalar potential which we take to be zero (up to small loop-induced effects). Once a specific UV model is considered in full detail, this assumption can be modified, and other consequences of it (such as corrections to the oblique $T$ parameter) should be considered.

As concerns the flavor structure of the fundamental couplings, we impose a global $U(2)_Q$ symmetry, under which the light left-handed $Q_{1,2}$ quarks transform as a doublet. This choice is taken to avoid large production rate at the LHC and dangerous FCNC transitions in the first and second generation\footnote{This typically suppresses the non-universality effect in $\psi(2S)$ decays to the permil level, and might not be necessary in some cases~\cite{U2Work}.}.
To determine uniquely the flavor structure of the NP couplings one should further specify the mass basis alignment of the UV operators. In what follows, we always consider alignment to the down mass basis, with $Q_3=\left(V_{u_ib}^*u_{L_i},b_L\right)$.
For the $\bar L_3L_3\bar Q_3Q_3$ and $\bar e_3L_3\bar Q_3d_3$ operators this choice is essential to ensure that $b\to c$ transition is modified, once $U(2)_Q$ is preserved. For the $\bar L_3e_3\bar Q_3u_2$ operators (generated by the $S$ and $D$ fields) one could choose alignment to the up-mass basis. In this case neither $\overline{c_L}c_R\to\tau^+\tau^-$ nor $\overline{b_L}b_R\to\tau^+\tau^-$ transitions are generated. Nevertheless, other NC operators which result in $\overline{b_L}b_L\to\tau^+\tau^-$ and $\overline{c_R}c_R\to\tau^+\tau^-$ transitions are formed, with the same CKM suppression (since $V_{tb}\simeq1$) and the same Wilson coefficient as in the down-aligned scenarios. We therefore find no change in our results once up-alignment is taken.

We assume no significant mixing between the NP and SM fields (this is crucial for the $W^\prime$ scenario), and take the quartic $\abs{X_{\rm NP}}^2\abs{H}^2$ couplings to be negligible.
To keep our models in the perturbative regime, we take all parameters to be smaller than $4\pi$ at the TeV scale. We stress again that, once an anomaly is found in LFU of $\Upsilon$ or $\psi$ decays, these assumptions should be reconsidered and a complete UV theory should be studied in full detail. Since we do not study the 2HDM case, condensation of the NP fields is absent due to Lorentz and/or $SU(3)_C$ symmetries. In the following, we take all the couplings to be real and denote $X_{ij}\equiv X_iX_j$.

To find the best fit values of the Wilson coefficients we minimize the $\chi^2$ function.
We use the experimental results quoted in Eq.~\eqref{eq:expR}. The theoretical uncertainties on the form factors which affect $R(D^{(*)})$ in the SM are small compared to the experimental errors, as evident from the accurate SM predictions. (See also Figs.~[1-2] in Ref~\cite{Bardhan:2016uhr}.) As for the other form factor ($F_T$ in the notations of Ref.~\cite{Bardhan:2016uhr}), we have explicitly checked that varying it within $20\%$ error does not alter our results for $R_{\tau/\ell}^V$. When considering a model, we are agnostic about how plausible it is from a model building point of view, and only compare it to the SM point. The latter gives $\chi^2\simeq 20$.

As concerns kinematical observables, the $q^2$ distribution of $\Gamma\left[B\to D^*\tau\nu\right]$ is hardly modified in all of the scenarios. We comment below on the $q^2$ distributions of $\Gamma\left[B\to D\tau\nu\right]$, which is, in general, consistent within the current uncertainties. Interference corrections, as analysed in Ref.~\cite{Ligeti:2016npd} are expected to be small.

\subsection{$W^\prime_\mu\sim(1,3)_0$}
We introduce a vector-boson, color-singlet, $SU(2)_L$-triplet $W^\prime_\mu\sim(1,3)_0$ with the following  interaction Lagrangian:
\beqa
\mathcal L_W&&=g_1\bar Q_3\tau_a\slashed{W}^aQ_3+g_2\bar L_3\tau_a\slashed{W}^aL_3.
\eeqa
Integrating out $W^\prime_\mu$, we obtain the following EFT Lagrangian:
\beqa
\mathcal L_W^{\rm EFT}&&=-\frac{g_1g_2}{M_{W^\prime}^2}\left(\bar Q_3\tau^a\gamma_\mu Q_3\right)\left(\bar L_3\tau_a\gamma^\mu L_3\right)=
-\frac{g_1g_2}{M_{W^\prime}^2}\mathcal{O}_{V_L}^2.
\eeqa

The relevant CC interactions are given by
\beqa
\mathcal L_{\rm CC}&&=-\frac{g_1g_2V_{cb}}{2M_{W^\prime}^2}\left(\bar \tau_L\gamma^\mu\nu_L\right)\left(\bar c_L\gamma_\mu b_L\right)+{\rm h.c.}.
\eeqa
The best-fit-point (BFP) and $95\%$ C.L. intervals are given by
\beqa
g_{12}^{\rm BFP}&&=7.7\left(\frac{M_{W^\prime}}{\rm TeV}\right)^2{\rm ~~with~~}
\chi^2=0.5,\no\\
g_{12}&&=\left[4.4,10.9\right]\left(\frac{M_{W^\prime}}{\rm TeV}\right)^2\,@\,95\%\,{\rm C.L.}.
\eeqa
where $g_{12} \equiv g_1 g_2$.
The $q^2$ distribution is identical to the SM one, since the new CC operator has the same Lorentz structure as in the SM.

The relevant NC interactions are given by
\beqa
\mathcal L_{\rm NC}&&=
\frac{V_{cb}^2g_1g_2}{4M_{W^\prime}^2}\left(\bar \tau_L\gamma^\mu \tau_L\right)\left(\bar c_L\gamma_\mu c_L\right)
-\frac{g_1g_2}{4M_W^2}\left(\bar \tau_L\gamma^\mu \tau_L\right)\left(\bar b_L\gamma_\mu b_L\right).
\eeqa
They induce both $\Upsilon\to\tau\tau$ and $\psi\to\tau\tau$.
Given the $95\%$ C.L. intervals quoted above, we find the following non-universalities
\beqa
R_{\tau/\ell}^{\Upsilon(1S)}&&=0.989-0.991,\no\\
R_{\tau/\ell}^{\Upsilon(2S)}&&=0.990-0.992,\no\\
R_{\tau/\ell}^{\Upsilon(3S)}&&=0.990-0.993,\no\\
R_{\tau/\ell}^{\psi(2S)}&&=0.390.
\eeqa

\subsection{$U_\mu\sim(3,1)_{+2/3}$}

We introduce a vector-boson, color-triplet, $SU(2)_L$-singlet $U_\mu\sim(3,1)_{+2/3}$ with the following  interaction Lagrangian:
\beqa
\mathcal L_U&&=g_1\bar Q_3\slashed{U}L_3+g_2\bar d_3\slashed{U}e_3+{\rm h.c.}.
\eeqa
Integrating out $U_\mu$, we obtain the following EFT Lagrangian:
\beqa
\mathcal L_U^{\rm EFT}&&=-\frac{\abs{g_1}^2}{M_U^2}\left(\bar Q_3\gamma_\mu L_3\right)\left(\bar L_3\gamma^\mu Q_3\right)
-\frac{\abs{g_2}^2}{M_U^2}\left(\bar e_3\gamma_\mu d_3\right)\left(\bar d_3\gamma^\mu e_3\right)-\left[\frac{g_1g_2^*}{M_U^2}\left(\bar Q_3\gamma^\mu L_3\right)\left(\bar e_3\gamma_\mu d_3\right)+{\rm h.c.}\right]\no\\
&&=
-\frac{\abs{g_1}^2}{2M_U^2}\mathcal{O}_{V_L}^1
-\frac{2\abs{g_1}^2}{M_U^2}\mathcal{O}_{V_L}^2
-\frac{\abs{g_2}^2}{M_U^2}\mathcal{O}_{V_R}^3
+\left[\frac{2g_1g_2^*}{M_U^2}\mathcal{O}_{S_R}+{\rm h.c.}\right].
\eeqa

\begin{figure}[h!]
	\begin{center}
	\includegraphics[height=2in]{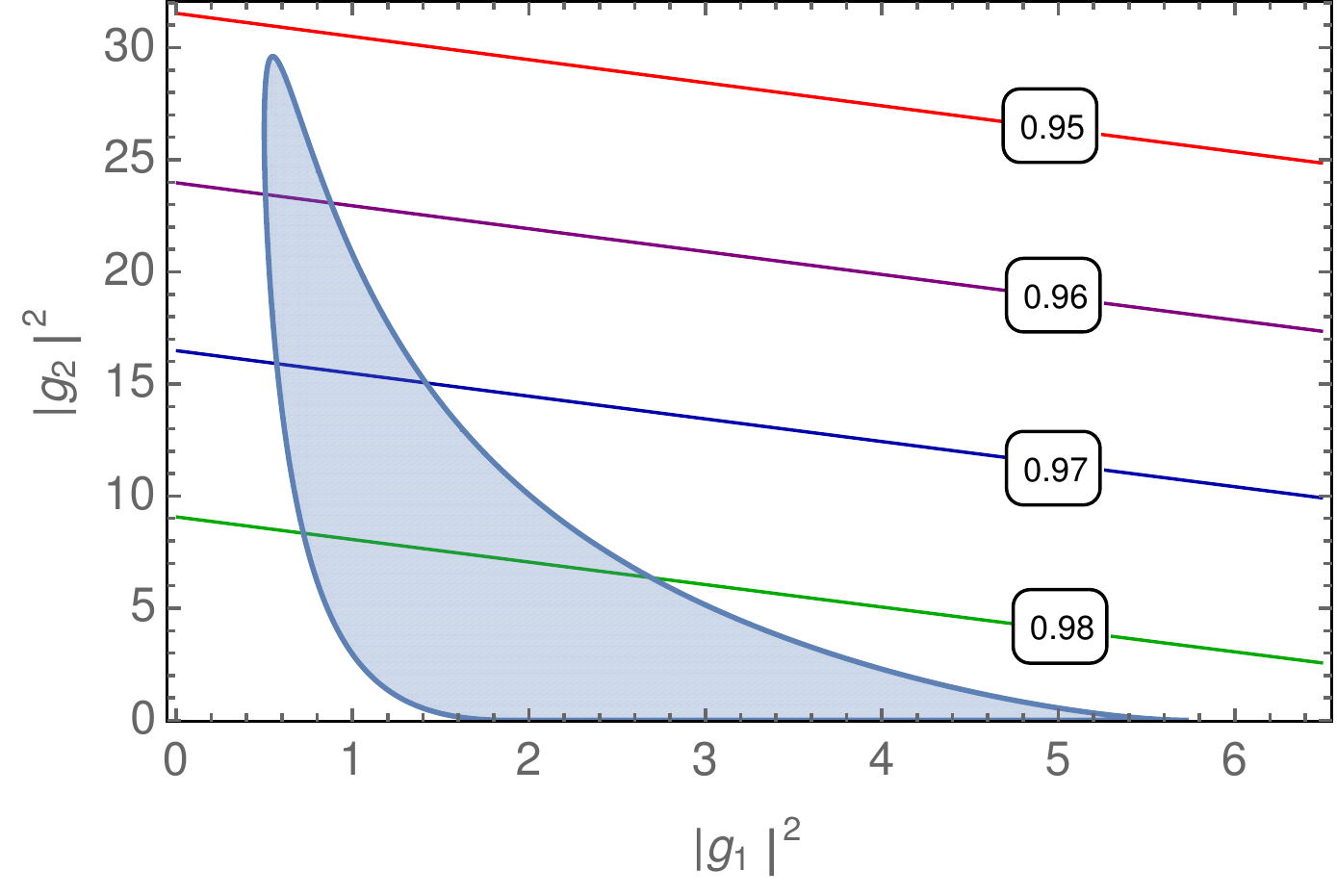}
\caption{\it $U_\mu\sim(3,1)_{2/3}$ couplings at $95\%$ C.L. for $M_U=1$ TeV. Colored lines are contours of $R_{\tau/\ell}^{\Upsilon(1S)}$.}
\label{fig:U_plot}
	\end{center}
\end{figure}

The relevant CC interactions are given by
\beqa
\mathcal L_{\rm CC}&&=
-\frac{\abs{g_1}^2V_{cb}}{M_U^2}\left(\bar \tau_L\gamma^\mu\nu_L\right)\left(\bar c_L\gamma_\mu b_L\right)
+\frac{2V_{cb}g_1g_2^*}{M_U^2}\left(\bar \tau_R \nu_L\right)\left(\bar c_L b_R\right)+{\rm h.c.}.
\eeqa
The BFP which explains $R(D^{(*)})$ is given by $g_1g_2<0$ and
\beqa
\left(\abs{g_1}^{2\,{\rm BFP}},\abs{g_{2}}^{2\,{\rm BFP}}\right)&&=\left(3.3,0.4\right)\left(\frac{M_U}{\rm TeV}\right)^2{\rm ~~with~~}
\chi^2=0.
\eeqa
The $95\%$ C.L. intervals are presented in Figure~\ref{fig:U_plot}.
The $q^2$ distribution of $\Gamma\left[B\to D\tau\nu\right]$ is modified compared to the SM one. Yet, as is evident from Figure.~\ref{fig:D_Umu}, this change is not very significant given the current uncertainties.

The relevant NC interactions are given by
\beqa
\mathcal L_{\rm NC}&&=
-\frac{\abs{g_1}^2}{M_U^2}\left(\bar \tau_L\gamma^\mu \tau_L\right)\left(\bar b_L\gamma_\mu b_L\right)
-\frac{\abs{g_2}^2}{M_U^2}\left(\bar \tau_R\gamma^\mu \tau_R\right)\left(\bar b_R\gamma_\mu b_R\right)
+\left[\frac{2g_1g_2^*}{M_U^2}\left(\bar \tau_R \tau_L\right)\left(\bar b_L b_R\right)+{\rm h.c.}\right].
\eeqa
They induce only $\Upsilon\to\tau\tau$.
Given the $95\%$ C.L. intervals quoted above, we find the following non-universalities
\beqa
R_{\tau/\ell}^{\Upsilon(1S)}&&=0.952-0.990,\no\\
R_{\tau/\ell}^{\Upsilon(2S)}&&=0.949-0.991,\no\\
R_{\tau/\ell}^{\Upsilon(3S)}&&=0.946-0.992.
\eeqa

\subsection{$X_\mu\sim(3,3)_{+2/3}$}

We introduce a vector-boson, color-triplet, $SU(2)_L$-triplet $X_\mu\sim(3,3)_{+2/3}$ with the following  interaction Lagrangian:
\beqa
\mathcal L_X&&=g\bar Q_3\tau_a\slashed{X}^aL_3+{\rm h.c.}.
\eeqa
Integrating out $X_\mu$, we obtain the following EFT Lagrangian:
\beqa
\mathcal L_X^{\rm EFT}&&=-\frac{\abs{g}^2}{M_X^2}\left(\bar Q_3\tau^a\gamma_\mu L_3\right)\left(\bar L_3\tau_a\gamma^\mu Q_3\right)=
-\frac{3\abs{g}^2}{8M_X^2}\mathcal{O}_{V_L}^1
+\frac{\abs{g}^2}{2M_X^2}\mathcal{O}_{V_L}^2.
\eeqa

The relevant CC interactions are given by
\beqa
\mathcal L_{\rm CC}&&=\frac{V_{cb}\abs{g}^2}{4M_X^2}\left(\bar \tau_L\gamma^\mu\nu_L\right)\left(\bar c_L\gamma_\mu b_L\right)+{\rm h.c.}.
\eeqa
The BFP and $95\%$ C.L. interval are given by
\beqa
\abs{g}^{2\,{\rm BFP}}&&=0\left(\frac{M_X}{\rm TeV}\right)^2{\rm ~~with~~}
\chi^2=20.4,\no\\
\abs{g}^2&&=\left[0,1.5\right]\left(\frac{M_X}{\rm TeV}\right)^2\,@\,95\%\,{\rm C.L.}.
\eeqa
The $q^2$ distribution is identical to the SM one, since the new CC operator has the same Lorentz structure as in the SM.

The relevant NC interactions are given by
\beqa
\mathcal L_{\rm NC}&&=
-\frac{V_{cb}^2\abs{g}^2}{2M_U^2}\left(\bar \tau_L\gamma^\mu \tau_L\right)\left(\bar c_L\gamma_\mu c_L\right)
-\frac{\abs{g}^2}{4M_U^2}\left(\bar \tau_L\gamma^\mu \tau_L\right)\left(\bar b_L\gamma_\mu b_L\right).
\eeqa
They induce both $\Upsilon\to\tau\tau$ and $\psi\to\tau\tau$. Given the $95\%$ C.L. intervals quoted above, we find the following non-universalities
\beqa
R_{\tau/\ell}^{\Upsilon(1S)}&&=0.992,\no\\
R_{\tau/\ell}^{\Upsilon(2S)}&&=0.993-0.994,\no\\
R_{\tau/\ell}^{\Upsilon(3S)}&&=0.994-0.995,\no\\
R_{\tau/\ell}^{\psi(2S)}&&=0.390.
\eeqa

\subsection{$S\sim(3,1)_{-1/3}$}

We introduce a scalar-boson, color-triplet, $SU(2)_L$-singlet $S\sim(3,1)_{-1/3}$ with the following  interaction Lagrangian:
\beqa
\mathcal{L}_S=&&\lambda_1S\bar L_3\epsilon^TQ_3^c+\lambda_2S\bar e_3u_2^c
+{\rm h.c.}.
\eeqa
We impose a global $3B-L$ symmetry, which prevent an additional Yukawa couplings of the form $Sdu$ and $SQQ$.
Integrating out $S$, we obtain the following EFT Lagrangian:
\beqa
\mathcal{L}_S^{\rm EFT}=&&\frac{\abs{\lambda_1}^2}{M_S^2}\abs{\bar L_3\epsilon^T Q_3^c}^2+\frac{\abs{\lambda_2}}{M_S^2}^2\abs{\bar e_3u_2^c}^2-\left[\frac{\lambda_1^*\lambda_2}{M_S^2}\left(\overline{L_3^c}\epsilon Q_3\right)\left(\bar e_3u_2^c\right)+{\rm h.c.}\right]\no\\
=&&
\frac{\abs{\lambda_1}^2}{4M_S^2}\mathcal{O}_{V_L}^1-\frac{\abs{\lambda_1}^2}{M_S^2}\mathcal{O}_{V_L}^2+\frac{\abs{\lambda_2}^2}{2M_S^2}\mathcal{O}_{V_R}^2+\left[\frac{\lambda_1^*\lambda_2}{2M_S^2}\mathcal{O}_{S_L}-\frac{\lambda_1^*\lambda_2}{8M_S^2}\mathcal{O}_T+{\rm h.c.}\right].
\eeqa

\begin{figure}[h!]
	\begin{center}
	\includegraphics[height=2in]{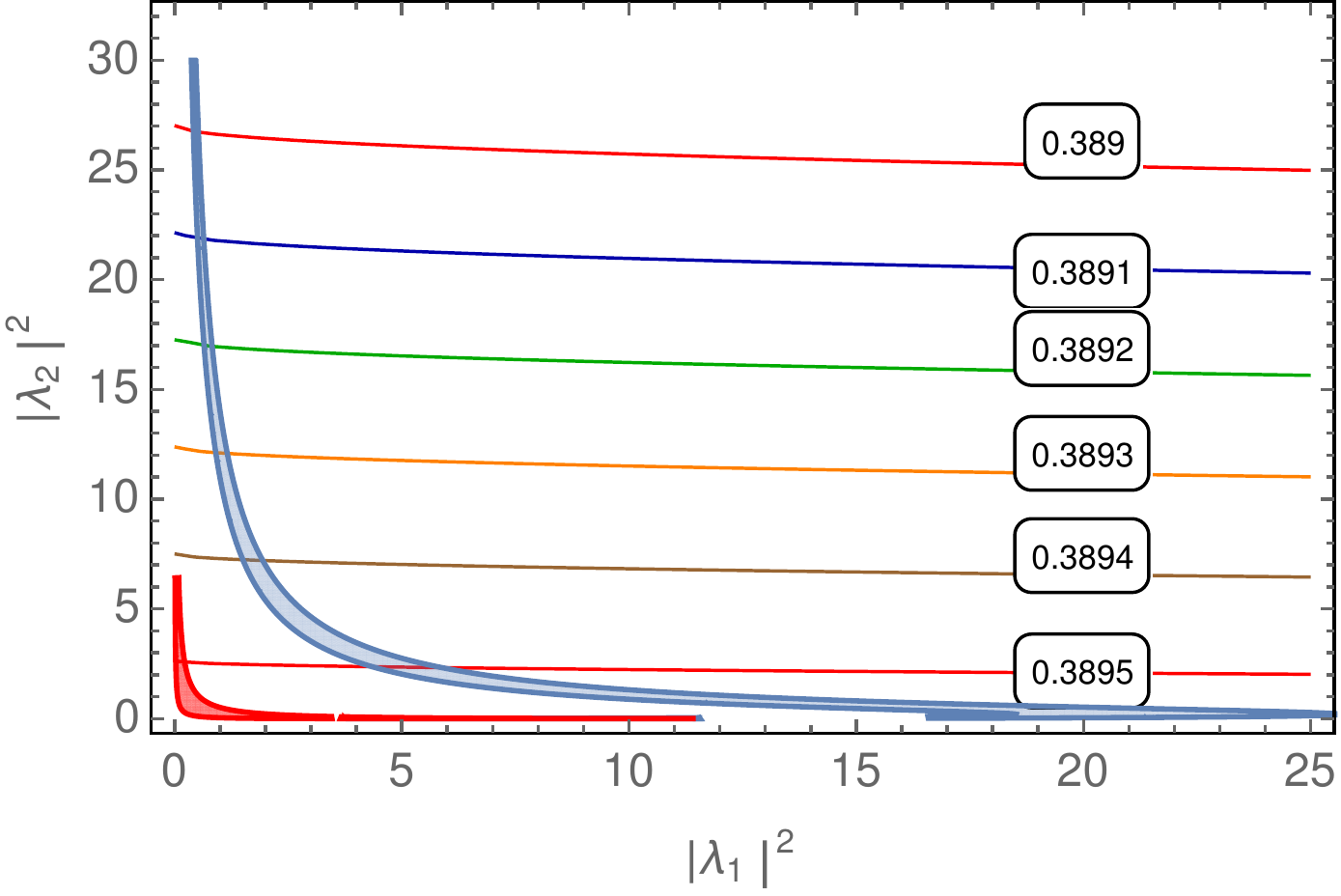}
\caption{\it $S\sim(3,1)_{-1/3}$ couplings at $95\%$ C.L. for $M_S=1$ TeV. Red region corresponds to $\lambda_1\lambda_2<0$ and blue region corresponds to $\lambda_1\lambda_2>0$.
Colored lines are contours of $R_{\tau/\ell}^{\psi(2S)}$}
\label{fig:S_plot}
	\end{center}
\end{figure}

The relevant CC interactions are given by
\beqa
\mathcal L_{\rm CC}&&=
-\frac{\abs{\lambda_1}^2V_{cb}}{2M_S^2}\left(\bar\tau_L\gamma^\mu\nu_L\right)\left(\bar c_L\gamma_\mu b_L\right)
+\frac{\lambda_1^*\lambda_2}{2M_S^2}\left(\bar\tau_R \nu_L\right)\left(\bar c_R b_L\right)-\frac{\lambda_1^*\lambda_2}{8M_S^2}\left(\bar\tau_R\sigma^{\mu\nu}\nu_L\right)\left(\bar c_R\sigma_{\mu\nu}b_L\right)+{\rm h.c.}.
\eeqa
The BFP which explains $R(D^{(*)})$ is given by $\lambda_1\lambda_2<0$ and
\beqa
\left(\abs{\lambda_1}^{2\,{\rm BFP}},\abs{\lambda_{2}}^{{2\,\rm BFP}}\right)&&=\left(0.3,0.9\right)\left(\frac{M_S}{\rm TeV}\right)^2{\rm ~~with~~}
\chi^2=0.
\eeqa
The $95\%$ C.L. intervals are presented in Figure~\ref{fig:S_plot}. The $q^2$ distribution of $\Gamma\left[B\to D\tau\nu\right]$ is modified compared to the SM one. Yet, as is evident from Figure.~\ref{fig:D_S}, this change is not very significant given the current uncertainties.

The relevant NC interactions are given by
\beqa
\mathcal L_{\rm NC}&&=
\frac{\abs{\lambda_1}^2}{2M_S^2}V_{cb}^2\left(\bar \tau_L\gamma^\mu \tau_L\right)\left(\bar c_L\gamma_\mu c_L\right)
+\frac{\abs{\lambda_2}^2}{2M_S^2}\left(\bar \tau_R\gamma^\mu \tau_R\right)\left(\bar c_R\gamma_\mu c_R\right)
\no\\
&&
+\left[-\frac{\lambda_1^*\lambda_2}{2M_S^2}V_{cb}\left(\bar \tau_R \tau_L\right)\left(\bar c_R c_L\right)
+\frac{\lambda_1^*\lambda_2}{8M_S^2}V_{cb}\left(\bar \tau_R\sigma^{\mu\nu}\tau_L\right)\left(\bar c_R\sigma_{\mu\nu}c_L\right)+{\rm h.c.}\right].
\eeqa
They induce only $\psi\to\tau\tau$. Given the $95\%$ C.L. intervals quoted above, we find the following non-universalities
\beqa
R_{\tau/\ell}^{\psi(2S)}&&=0.389-0.390.
\eeqa

\subsection{$T\sim(3,3)_{-1/3}$}
We introduce a scalar-boson, color-triplet, $SU(2)_L$-triplet $S\sim(3,3)_{-1/3}$ with the following  interaction Lagrangian:
\beqa
\mathcal{L}_T=&&\lambda T^a\bar L_3\tau_a\epsilon^TQ_3^c
+\lambda^*T^{*a}\overline{Q_3^c}\epsilon\tau_a L_3.
\eeqa
We impose global $3B-L$ symmetry to forbid $TQQ$ terms.
Integrating out $T$, we obtain the following EFT Lagrangian:
\beqa
\mathcal{L}_T^{\rm EFT}=&&\frac{1}{M_T^2}\abs{\lambda}^2\left(\bar L_3\tau_a\epsilon^TQ_3^c\right)\left(\overline{Q_3^c}\epsilon\tau_a L_3\right)=\frac{3\abs{\lambda}^2}{16M_T^2}\mathcal{O}_{V_L}^1+\frac{\abs{\lambda}^2}{4M_T^2}\mathcal{O}_{V_L}^2.
\eeqa

The relevant CC interactions are given by
\beqa
\mathcal L_{\rm CC}&&=\frac{\abs{\lambda}^2V_{cb}}{8M_T^2}\left(\bar \tau_L\gamma^\mu\nu_L\right)\left(\bar c_L\gamma_\mu b_L\right).
\eeqa
The BFP and $95\%$ C.L. interval are given by
\beqa
\abs{\lambda}^{2\,{\rm BFP}}&&=0\left(\frac{M_T}{\rm TeV}\right)^2{\rm ~~with~~}
\chi^2=20.4,\no\\
\abs{\lambda}^2&&=\left[0,3.0\right]\left(\frac{M_T}{\rm TeV}\right)^2\,@\,95\%\,{\rm C.L.}.
\eeqa
The $q^2$ distribution is identical to the SM one, since the new CC operator has the same Lorentz structure as in the SM.

The relevant NC interactions are given by
\beqa
\mathcal L_{\rm NC}&&=\frac{\abs{\lambda}^2V_{cb}^2}{8M_T^2}\left(\bar \tau_L\gamma^\mu \tau_L\right)\left(\bar c_L\gamma_\mu c_L\right)+
\frac{\abs{\lambda}^2}{4M_T^2}\left(\bar \tau_L\gamma^\mu \tau_L\right)\left(\bar b_L\gamma_\mu b_L\right).
\eeqa
They induce both $\Upsilon\to\tau\tau$ and $\psi\to\tau\tau$. Given the $95\%$ C.L. intervals quoted above, we find the following non-universalities
\beqa
R_{\tau/\ell}^{\Upsilon(1S)}&&=0.992-0.993,\no\\
R_{\tau/\ell}^{\Upsilon(2S)}&&=0.994-0.995,\no\\
R_{\tau/\ell}^{\Upsilon(3S)}&&=0.995-0.996,\no\\
R_{\tau/\ell}^{\psi(2S)}&&=0.390.
\eeqa

\subsection{$\phi\sim(1,2)_{+1/2}$}
A scalar-boson, color-singlet, $SU(2)_L$-doublet $\phi\sim(1,2)_{+1/2}$ does not affect $V\to\ell\ell$ as it generates only scalar couplings.

\subsection{$D\sim(3,2)_{+7/6}$}
We introduce a scalar-boson, color-triplet, $SU(2)_L$-doublet $D\sim(3,2)_{+7/6}$ with the following  interaction Lagrangian:
\beqa
\mathcal{L}_D=&&\lambda_1D\bar Q_3e_3+\lambda_2D\epsilon \bar u_2L_3+{\rm h.c.}.
\eeqa
Integrating out $D$, we obtain the following EFT Lagrangian:
\beqa
\mathcal{L}_D^{\rm EFT}&&=\frac{\abs{\lambda_1}^2}{M_D^2}\abs{\bar Q_3e_3}^2-\frac{\abs{\lambda_2}^2}{M_D^2}\abs{\bar u_2L_3}^2-\left[\frac{\lambda_1^*\lambda_2}{M_D^2}\left(\bar u_2L_3\right)\epsilon\left(\bar e_3Q_3\right)\right]\no\\
&&=-\frac{\abs{\lambda_1}^2}{2M_D^2}\mathcal{O}_{V_R}^1+\frac{\abs{\lambda_2}^2}{2M_D^2}\mathcal{O}_{V_L}^3+
\left[\frac{\lambda_1^*\lambda_2}{2M_D^2}\left(\mathcal{O}_{S_L}+\frac{1}{4}\mathcal{O}_{T}\right)+{\rm h.c.}\right].
\eeqa

The relevant CC interactions are given by
\beqa
\mathcal L_{\rm CC}&&=
\frac{\lambda_1^*\lambda_2}{2M_D^2}\left(\bar \tau_R \nu_L\right)\left(\bar c_R b_L\right)
+\frac{\lambda_1^*\lambda_2}{8M_D^2}\left(\bar \tau_R\sigma^{\mu\nu}\nu_L\right)\left(\bar c_R\sigma_{\mu\nu}b_L\right)+{\rm h.c.}.
\eeqa
The BFP and $95\%$ C.L. interval are given by
\beqa
\lambda_{12}^{\rm BFP}&&=0.3\left(\frac{M_D}{\rm TeV}\right)^2{\rm ~~with~~}
\chi^2=17,\no\\
\lambda_{12}&&=\left[0,0.6\right]\left(\frac{M_D}{\rm TeV}\right)^2\,@\,95\%\,{\rm C.L.}.
\eeqa
The $q^2$ distribution of $\Gamma\left[B\to D\tau\nu\right]$ is modified compared to the SM one. Yet, as is evident from Figure.~\ref{fig:D_D}, this change is not very significant given the current uncertainties.

The relevant NC interactions are given by
\beqa
\mathcal L_{\rm NC}=&&
-\frac{\abs{\lambda_1}^2}{2M_D^2}V_{cb}^2\left(\bar \tau_R\gamma^\mu \tau_R\right)\left(\bar c_L\gamma_\mu c_L\right)
-\frac{\abs{\lambda_1}^2}{2M_D^2}\left(\bar \tau_R\gamma^\mu \tau_R\right)\left(\bar b_L\gamma_\mu b_L\right)
+\frac{\abs{\lambda_2}^2}{2M_D^2}\left(\bar \tau_L\gamma^\mu \tau_L\right)\left(\bar c_R\gamma_\mu c_R\right) \no\\
&& -\left[\frac{\lambda_1^*\lambda_2}{2 M_D^2}V_{cb}
\left(\bar\tau_R\tau_L\right)\left(\bar c_R c_L\right)
+\frac{\lambda_1^*\lambda_2}{8 M_D^2}V_{cb}\left(\bar\tau_R\sigma^{\mu\nu}\tau_L\right)\left(\bar c_R\sigma_{\mu\nu}c_L\right)+{\rm h.c.}\right].
\eeqa
They induce both $\Upsilon\to\tau\tau$ and $\psi\to\tau\tau$. Given the $95\%$ C.L. intervals quoted above, we find the following non-universalities
\beqa
R_{\tau/\ell}^{\Upsilon(1S)}&&=0.889-0.992,\no\\
R_{\tau/\ell}^{\Upsilon(2S)}&&=0.879-0.994,\no\\
R_{\tau/\ell}^{\Upsilon(3S)}&&=0.873-0.995,\no\\
R_{\tau/\ell}^{\psi(2S)}&&=0.386-0.390.
\eeqa

\subsection{$V_\mu\sim(3,2)_{-5/6}$}
We introduce a vector-boson, color-triplet, $SU(2)_L$-doublet $V_\mu\sim(3,2)_{-5/6}$ with the following  interaction Lagrangian:
\beqa
\mathcal L_V&&=g_1\bar Q_3\slashed{V}e_3^c+g_2\bar L_3\slashed{V}d_3^c+{\rm h.c.}.
\eeqa
Integrating out $V_\mu$, we obtain the following EFT Lagrangian:
\beqa
\mathcal L_V^{\rm EFT}&&=-\frac{\abs{g_1}^2}{M_V^2}\left(\bar Q_3\gamma_\mu e_3^c\right)\left(\bar e_3^c\gamma^\mu Q_3\right)
-\frac{\abs{g_2}^2}{M_V^2}\left(\bar L_3\gamma_\mu d_3^c\right)\left(\bar d_3^c\gamma^\mu L_3\right)-\left[\frac{g_1g_2^*}{M_V^2}\left(\bar Q_3\gamma^\mu e_3^c\right)\left(\bar d_3^c\gamma_\mu L_3\right)+{\rm h.c.}\right]\no\\
&&=
-\frac{\abs{g_1}^2}{M_V^2}\mathcal{O}_{V_R}^1
-\frac{\abs{g_2}^2}{M_V^2}\mathcal{O}_{V_L}^4
-\left[\frac{2g_1g_2^*}{M_V^2}\mathcal{O}_{S_R}+{\rm h.c.}\right].
\eeqa

\begin{figure}[h!]
  \subfloat[ $~SM$]
  {\includegraphics[scale=0.5]
  	{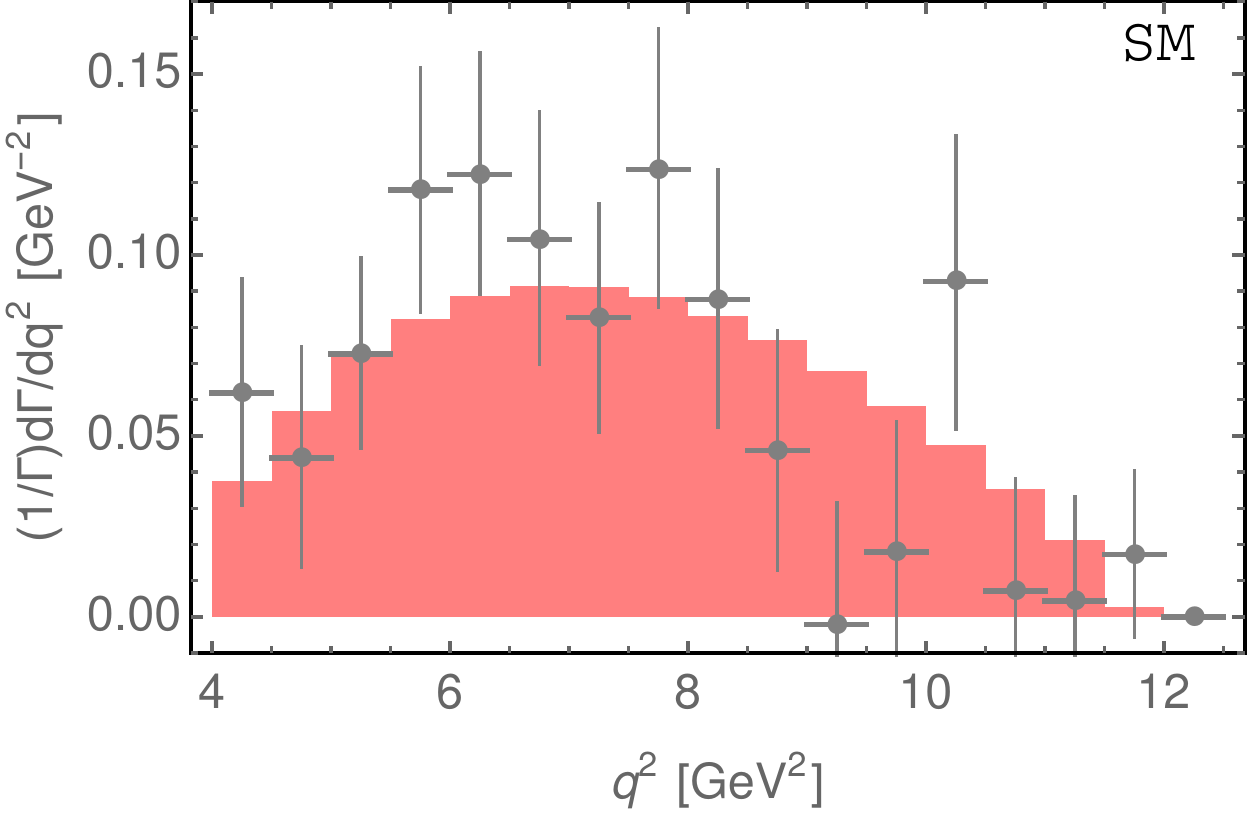}\label{fig:D_SM}}
  \quad
  \subfloat[ ~$U_\mu$]
  {\includegraphics[scale=0.5]
  	{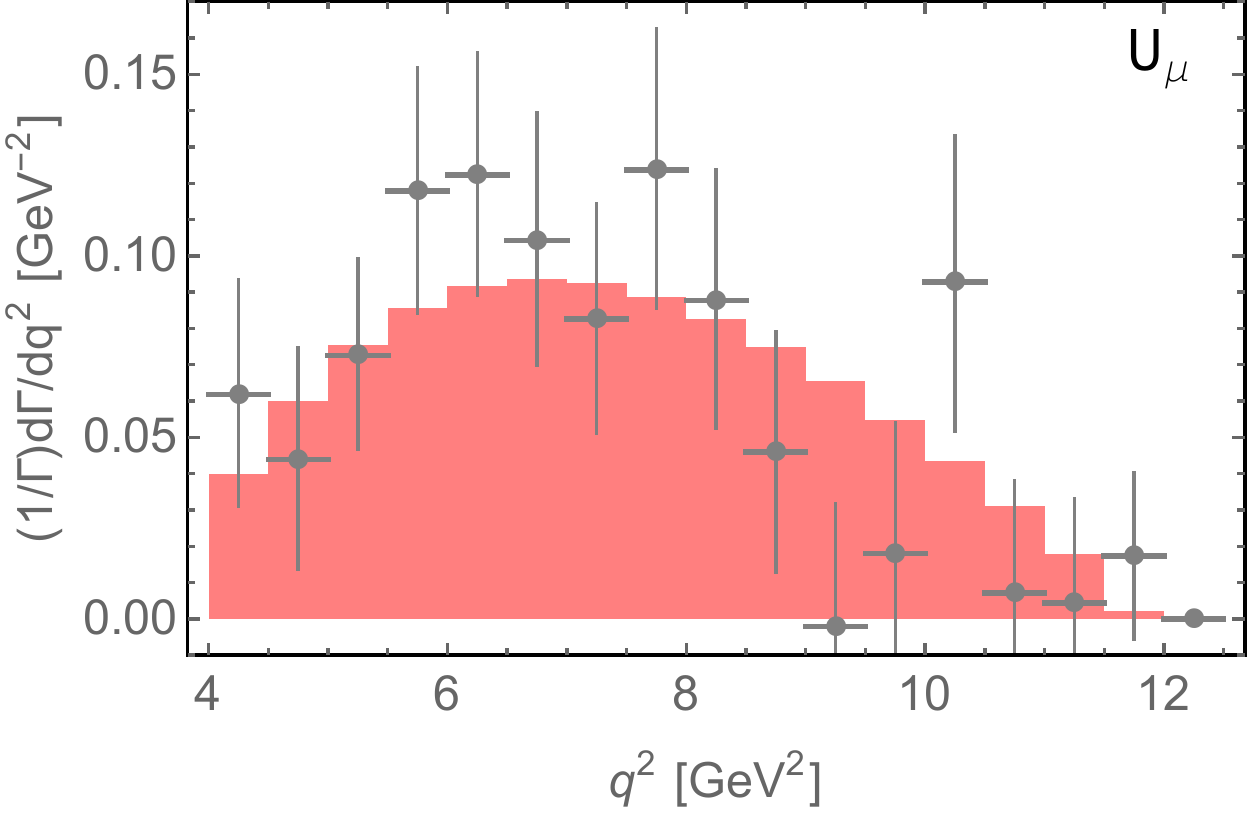}\label{fig:D_Umu}}
    \quad
  \subfloat[ ~$S$]
  {\includegraphics[scale=0.5]
  	{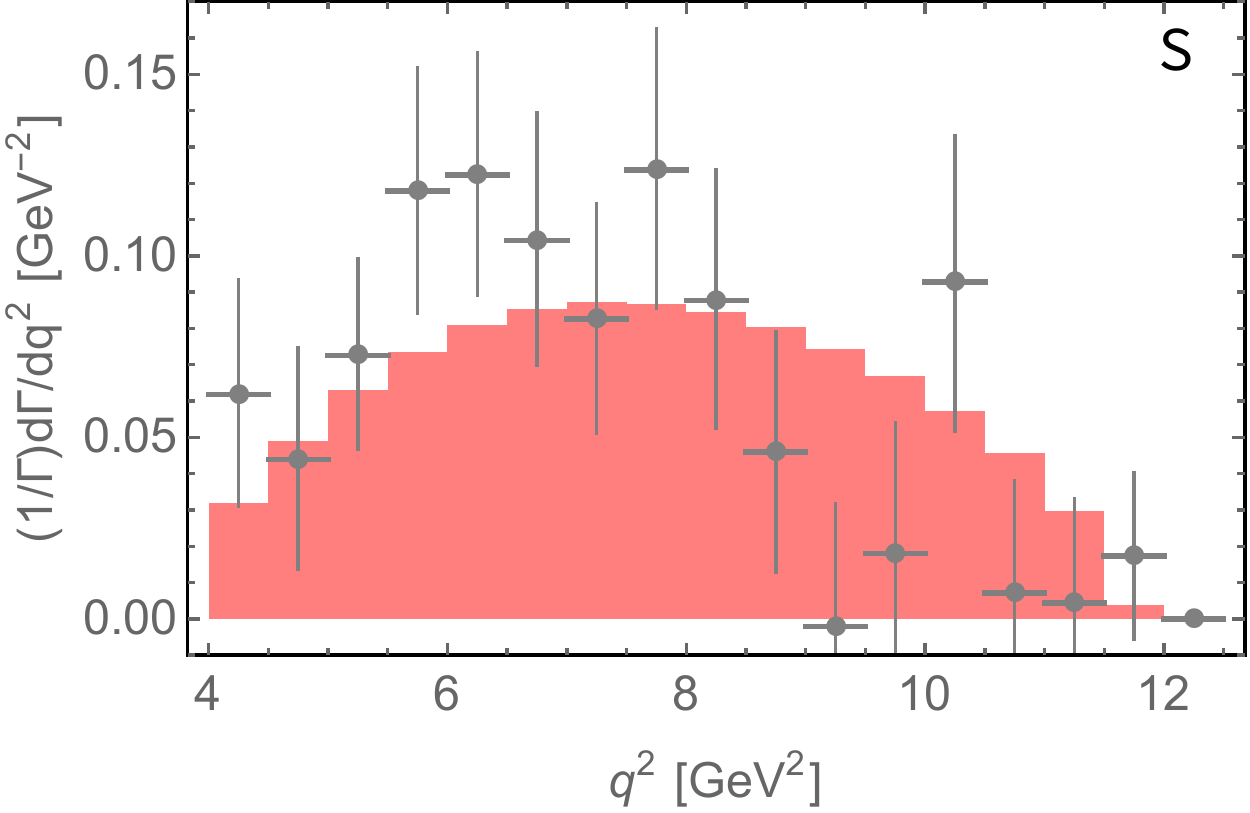}\label{fig:D_S}  }
      \quad
  \subfloat[ ~$D$]
  {\includegraphics[scale=0.5]
  	{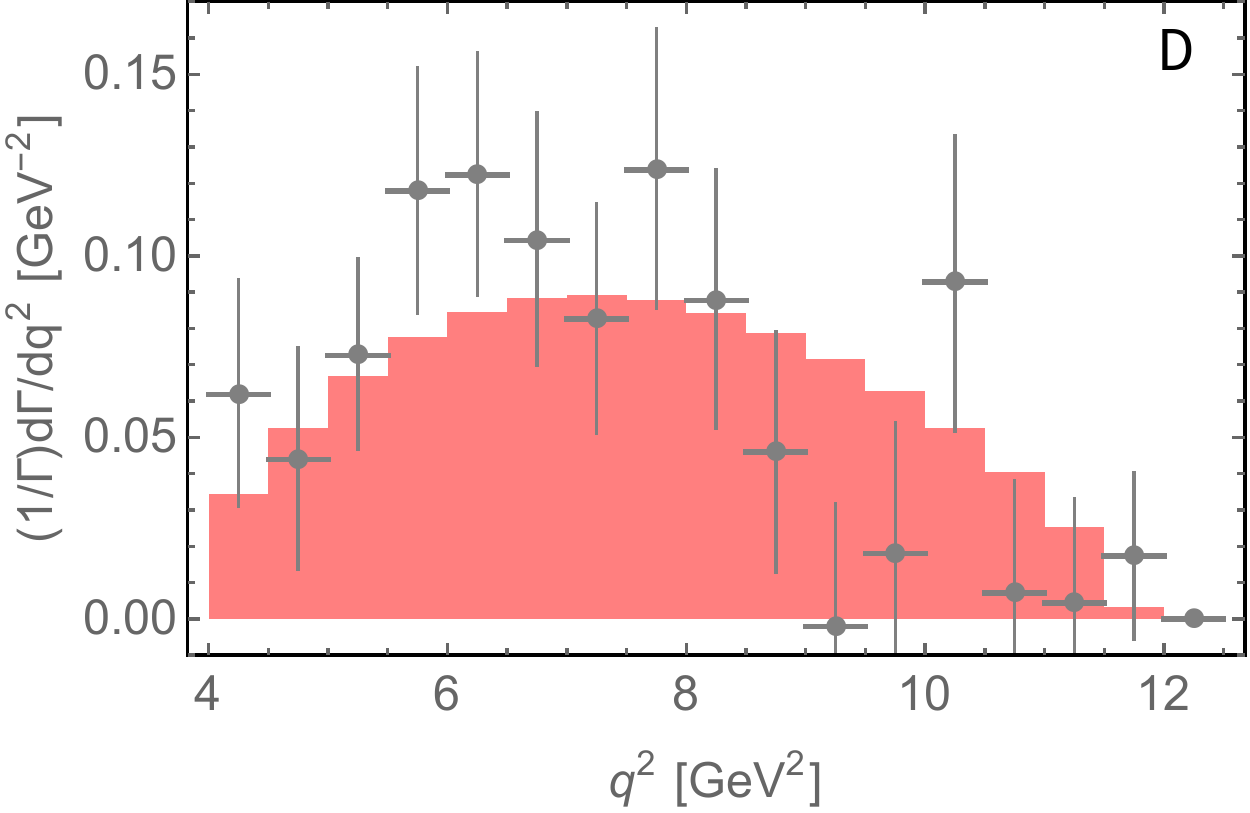}\label{fig:D_D}}
      \quad
  \subfloat[ ~$V_\mu$]
  {\includegraphics[scale=0.5]
  	{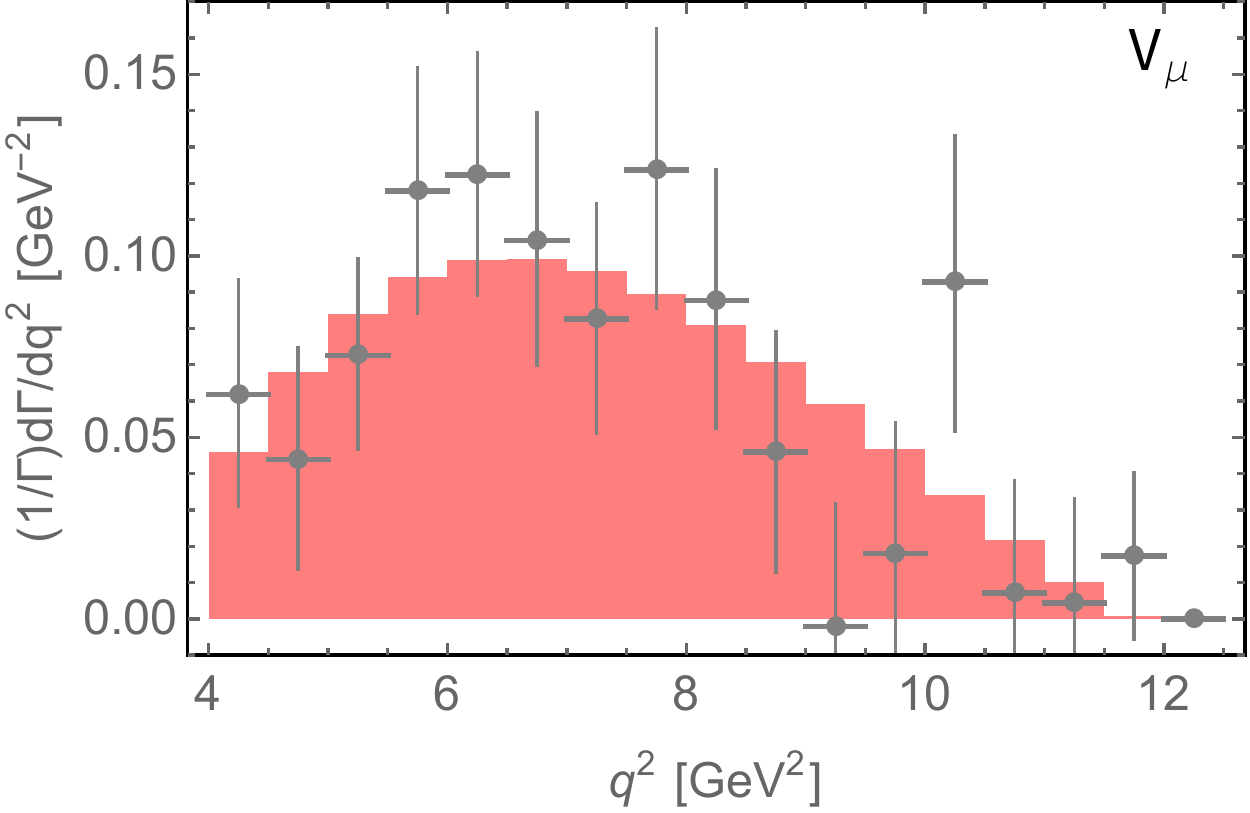}\label{fig:D_Vmu} }
  \caption{\it Normalized $q^2$ distribution of $\Gamma[B\to D\tau\nu]$. Data points and error bars are taken from Ref.~\cite{Lees:2013uzd}.}
 \label{fig:q2_D}
\end{figure}

The relevant CC interactions are given by
\beqa
\mathcal L_{\rm CC}&&=
-\frac{2V_{cb}g_1g_2^*}{M_V^2}\left(\bar \tau_R \nu_L\right)\left(\bar c_L b_R\right)+{\rm h.c.}.
\eeqa
The BFP and $95\%$ C.L. interval are given by
\beqa
g_{12}^{\rm BFP}&&=4.0\left(\frac{M_V}{\rm TeV}\right)^2{\rm ~~with~~}
\chi^2=8.2,\no\\
g_{12}&&=\left[1.9,5.9\right]\left(\frac{M_V}{\rm TeV}\right)^2\,@\,95\%\,{\rm C.L.}.
\eeqa
The $q^2$ distribution of $\Gamma\left[B\to D\tau\nu\right]$ is modified compared to the SM one. Yet, as is evident from Figure.~\ref{fig:D_Vmu}, this change is not very significant given the current uncertainties.

The relevant NC interactions are given by
\beqa
\mathcal L_{\rm NC}&&=
-\frac{\abs{g_1}^2}{M_V^2}\left(\bar \tau_L\gamma^\mu \tau_L\right)\left(\bar b_L\gamma_\mu b_L\right)
-\frac{\abs{g_2}^2}{M_V^2}\left(\bar \tau_R\gamma^\mu \tau_R\right)\left(\bar b_R\gamma_\mu b_R\right)
-\left[\frac{2g_1g_2^*}{M_V^2}\left(\bar \tau_R \tau_L\right)\left(\bar b_L b_R\right)+{\rm h.c.}\right].
\eeqa
They induce only $\Upsilon\to\tau\tau$. Given the $95\%$ C.L. intervals quoted above, we find the following non-universalities
\beqa
R_{\tau/\ell}^{\Upsilon(1S)}&&=0.976-0.987,\no\\
R_{\tau/\ell}^{\Upsilon(2S)}&&=0.976-0.988,\no\\
R_{\tau/\ell}^{\Upsilon(3S)}&&=0.976-0.988.
\eeqa

\section{Discussion and future prospects}\label{sec:prospects}
The allowed ranges for $R^V_{\tau/\ell}$ in simplified models that account for the deviations of $R(D^{(*)})$ are presented in Table \ref{tab:results}. As concerns $R^{\psi(2S)}_{\tau/\ell}$, it is modified by no more than three permil, while the present experimental accuracy is of order thirteen percent. Thus, the maximal modification is about a factor of fifty below current sensitivity. We conclude that $R^{\psi(2S)}_{\tau/\ell}$ does not probe at present models that solve the $R(D^{(*)})$ puzzle. (Removing the imposed $U(2)_Q$ symmetry might lead to a much larger modification of $R^{\psi(2S)}_{\tau/\ell}$~\cite{U2Work}.)

As concerns $R^{\Upsilon(1S)}_{\tau/\ell}$, in all simplified models, except the $T$ model, it is smaller than the SM. The lowest value is  $4\%$ below the SM value. The current experimental accuracy is $2.5\%$, comparable to the predicted deviations. Furthermore, the central value is higher (by about $0.5\sigma$) then the SM prediction. We conclude that $R^{\Upsilon(1S)}_{\tau/\ell}$ is starting to probe relevant models, disfavoring parts of the parameter space in some of the models.

Non-universality in leptonic $\Upsilon$ decays was tested by CLEO~\cite{Besson:2006gj} for the $1S,2S$ and $3S$ states, and by BaBar~\cite{delAmoSanchez:2010bt} for the $1S$ state. These measurements read
\beqa
R^{\Upsilon(1S),\,{\rm BaBar}}_{\tau/\mu}&&=1.005\pm0.013_{\rm stat}\pm0.022_{\rm syst},\no\\
R^{\Upsilon(1S),\,{\rm CLEO}}_{\tau/\mu}&&=1.02\pm0.02_{\rm stat}\pm0.05_{\rm syst},\no\\
R^{\Upsilon(2S),\,{\rm CLEO}}_{\tau/\mu}&&=1.04\pm0.04_{\rm stat}\pm0.05_{\rm syst},\no\\
R^{\Upsilon(3S),\,{\rm CLEO}}_{\tau/\mu}&&=1.05\pm0.08_{\rm stat}\pm0.05_{\rm syst}.
\eeqa
CLEO's data used in this analysis includes both on-resonance and off-resonance subsamples, which correspond approximately to $21,\ 10$, and $6$ million events of the $1S,\ 2S$ and $3S$ states, respectively.
The $\Upsilon(2S)$ on-resonance sample collected by BaBar (Belle) is about $\sim10\,(16)$ times larger than CLEO's sample. The $\Upsilon(3S)$ on-resonance sample collected by BaBar (Belle) is about $\sim20\,(2)$ times larger than CLEO's sample.
Analyzing these existing data sets will allow to reduce each of the statistical uncertainties to roughly $1-2$ percent. Additional improvement can be achieved by using the $\Upsilon$ cascade chains, which will render this error negligible.

The systematic error in BaBar's analysis is currently controlled by the uncertainties on the different $\tau$ and $\mu$ total efficiencies and event shapes.
The larger statistics can realistically lead to a reduction of this uncertainty by a factor of two.
Altogether, we estimate that a total uncertainty of about $1\%$ can be obtained by analyzing the existing data. Future measurement in Belle II can further reduce this uncertainty. Assuming that the important systematical error is governed by the limited statistics, we estimate that reaching a $\sigma_{\rm sys}=0.4\%$ for $R^{\Upsilon(1S),\,{\rm Belle~II}}_{\tau/\mu}$ would require integrated luminosity of $\mathcal{L}\sim1/{\rm ab}$ at the $\Upsilon(3S)$ energy. Thus, our study gives additional motivation to the proposal of Ref.~\cite{Ye:2016pgb} to study the $\Upsilon(3S)$ resonance at the early physics program of the Belle II experiment.

One important point to emphasize is that theoretically any model that affects an $\Upsilon$ state affects them all. Thus, it is wise to combine the experimental results of $\Upsilon(1S)$, $\Upsilon(2S)$, and $\Upsilon(3S)$ into one test of universality.
Eq.~\eqref{eq:R_QED} as a function of $m_{\Upsilon(nS)}$,
\begin{align}
R_{\tau/\ell}(m_{\Upsilon(nS)}) = \left[1+ 2x_{\tau,1S}^2 \left(\frac{m_{\Upsilon(1S)}}{m_{\Upsilon(nS)}}\right)^2\right]\left[1 -4x_{\tau,1S}^2 \left(\frac{m_{\Upsilon(1S)}}{m_{\Upsilon(nS)}}\right)^2\right]^{1/2}~,
\end{align}
where $x_{\tau,1S}\equiv m_\tau/m_{\Upsilon(1S)}=0.187823$, can be used to test the SM, and a violation of it would constitute a signal for NP.
The correlation of the systematic uncertainties between the
different $\Upsilon$ states is probably large (if not maximal), not
allowing to further reduce this part of the error with a combined
analysis.

BESII~\cite{Ablikim:2006iq} measurement reads ${\rm BR}(\psi(2S)\to\tau^+\tau^-)=\left(3.08\pm0.21_{\rm stat}\pm0.38_{\rm sys}\right)\times10^{-3}$ using 14M $\psi(2S)$ events. BESII already collected 106M events which will reduce the statistical error by a factor of $\sim3$. (KEDR also measures this tauonic branching ratio~\cite{Anashin:2007zz} but it is not used by the PDG fit.) 
 The relative systematic uncertainty on ${\rm BR}(\psi(2S)\to\mu^+\mu^-)$ (as measured by BaBar) is $10\%$~\cite{Aubert:2001ak}, while the relative systematic error on ${\rm BR}(\psi(2S)\to {e^+e^-})$ (as measured by BESII) is approximately $4\%$~\cite{Ablikim:2008zzb}. It is then not very likely that the ratio $R_{\tau/\ell}^{\psi(2S)}$ will be measured to an accuracy better than $4\%$. We conclude that at least an order of magnitude improvement in this uncertainty is needed to be achieved in Bess III to start probing the relevant parameter space.

\section{Summary and conclusions}\label{sec:concs}

There is a $3.9\sigma$ evidence that the ratio $R(D^{(*)})\equiv\Gamma(B\to D^{(*)}\tau\nu)/\Gamma(B\to D^{(*)}\ell\nu)$, where $\ell=\mu,e$, is considerably enhanced compared to the Standard Model value. To explain a large enhancement of a SM tree-level process, the required new physics is likely to include new bosons which mediate the $B\to D^{(*)}\tau\nu$ decay at tree level. There are seven such candidates, and none is a SM singlet. Thus, it is highly likely that their mass is well above $m_B$, which allows one to examine their effects in an EFT language. Specifically, they should generate dimension-six, four-fermi (two quark and two lepton fields) operators.

In the absence of light right-handed neutrinos, the four-fermi operators include one or two $SU(2)_L$-doublet quark fields. Consequently, a variety of processes, in addition to $B\to D^{(*)}\tau\nu$, are affected. Some of these, such as $t\to c\tau^+\tau^-$ decay, $B_c\to\tau\nu$ decay~\cite{Li:2016vvp,Alonso:2015sja}, $\Lambda_b\to\Lambda_c\tau\bar\nu_\tau$ decay~\cite{Li:2016pdv,lambda_b2}, and $b\bar b\ / c\bar c\to\tau^+\tau^-$ scattering~\cite{Faroughy:2016osc}, have been previously studied in the literature, and probe the proposed models. Here, we suggest another class of observables: lepton non-universality in leptonic decays of the $\Upsilon$ and $\psi$ vector-mesons, parameterized by the ratio $R^{V}_{\tau/\ell}\equiv\Gamma(V\to\tau\tau)/\Gamma(V\to\ell\ell)$.

We find that, once a $U(2)_Q$ symmetry is assumed, current measurements of $\psi$ decays do not have the accuracy required to probe the models in a significant way.  On the other hand, for $\Upsilon(1S)$ decays, the current experimental accuracy and the predicted deviations are of the same order of magnitude. A modest improvement in the experimental accuracy is capable of favoring some models and disfavoring others. If, for example, it is established that $R^{\Upsilon(1S)}_{\tau/\ell}$ is larger than the SM value, than all the simplified models will be disfavored. (The only model which enhances $R^{\Upsilon(1S)}_{\tau/\ell}$ is the $T$ model, which gives, however, a very small effect.) If experiments reach high accuracy in the leptonic vector meson decays and observe no signal, then the models that allow negligible deviations will be favored. Another possibility is that the light neutrino has a significant $\nu_R$ component, in which case $R(D^{(*)})$ could be explained by operators which do not affect $V\to\tau\tau$.

We discussed the prospects of such improvement in the experimental accuracy. Current data samples should already allow to reduce the statistical error considerably and reach an accuracy of about 1.5\%. If Belle II operates below the $\Upsilon(4S)$ resonance, it can contribute significantly, via the measurements of $R^{\Upsilon(nS)}_{\tau/\ell}$, to our understanding of the $R(D^{(*)})$ puzzle.

\acknowledgments
We thank Avital Dery, Admir Greljo, Avner Soffer, Alexander Penin and Nadav Priel for useful discussions. We
are thankful to Diptimoy Ghosh for his great help with the numerical
evaluation of $R(D^{(*)})$.
This research is supported by the United States-Israel Binational
Science Foundation (BSF), Jerusalem, Israel (grant number 2014230),
and by the I-CORE program of the Planning and Budgeting Committee and
the Israel Science Foundation (grant number 1937/12).
YG is supported in part by the NSF grant PHY-1316222.
YN is the Amos de-Shalit chair of theoretical physics.
YN is supported by grants from the Israel Science Foundation (grant
number 394/16) and from the Minerva Foundation.

\appendix

\section{The leptonic width in the SM}\label{app:SM}

The leading SM decay rate is given by the Van Royen-Weisskopf formula~\cite{VanRoyen:1967nq},
\beqa
\Gamma^0\left[V\to\ell\ell\right]&&=4\pi\alpha^2\left(\frac{f_V^2}{m_V}\right)Q_q^2(1-4x_\ell^2)^{1/2}\left(1+2x_\ell^2\right).
\eeqa
The dominant corrections arise from tree-level $Z$ exchange and quantum QED and QCD corrections,
\beqa
\Gamma\left[V\to\ell\ell\right]&&\simeq\Gamma^0\left[V\to\ell\ell\right]\left(1+\delta_Z^{\rm tree}+\delta_{\rm QCD}+\delta_{\rm EM}\right).
\eeqa
The tree-level $Z$ mediated correction is
\beqa
\delta_Z^{\rm tree}=
\left(\frac{\epsilon}{\epsilon-1}\right)\left(\frac{g_V^q}{4Q_qc_W^2s_W^2}\right)\left(1-4s_W^2\right)+\left(\frac{\epsilon}{\epsilon-1}\right)^2
\left(\frac{g_V^q}{8Q_qc_W^2s_W^2}\right)^2\left[\left(1-4s_W^2\right)^2+\left(\frac{1-4x_\ell^2}{1+2x_\ell^2}\right)\right].
\eeqa
Here $\epsilon=m_V^2/m_Z^2$ and $g_V^q=T_3^q-2Q_qs_W^2$. This is an $\mathcal{O}(10^{-4})$ correction, well below the experimental sensitivity. The resulting change in the non-universality relation (Eq.~\eqref{eq:R_QED}) is, however, only $\mathcal{O}(10^{-5})$, as the interference with the photon projects out the vectorial part of the $Z$ coupling. QCD corrections have been studied extensively in the literature (see, for example,~\cite{Beneke:2014qea} and references therein). A crucial point is that these short- and long-distance QCD effects do not depend on $m_\ell$, and cancel in the ratio $R^V_{\tau/\ell}$. These can be absorbed in the definition of the vector meson form factor, $f_V$.

The leading QED corrections which are not included in the definition of $f_V$ arise from corrections to the photon two-point function, the photon-lepton vertex corrections, and the irreducible real photon emission (FSR). Corrections due to two-photon exchange are absent: The Landau-Yang theorem~\cite{Landau:1948kw,Yang:1950rg} implies $\mathcal{M}(V\to\gamma\gamma)=0$ for massive vector bosons. Therefore exploiting the optical theorem and dispersion relation for real analytic functions, as well as QED and QCD CP invariance, forbids $\mathcal{M}(V\to\ell\ell)$ via two photon exchange. The corrections to $\Pi_{\gamma\gamma}$ are taken into account by using the running couplings, namely evaluating $\alpha(\mu)$ at $\mu=m_V$. Once again these are universal corrections that do not affect $R^V_{\tau/\ell}$.

As for the leptonic-vertex corrections, these exhibit IR singularities which are regulated by the real emission of soft photons. Clearly, the latter depends on the experimental resolution and should be determined (and unfolded) by the experiments using state-of-the-art detector simulation in MC analysis. Following the LEP report for $Z$ pole observables~\cite{ALEPH:2005ab}, we quote here the inclusive $\Gamma[V\to\ell^+\ell^-+\gamma]$ at one-loop order in QED, which does not depend on the experimental setup. It reads~\cite{Bardin:1999ak}
\beqa
\delta_{\rm EM}&&=\frac{\alpha}{4\pi}
\left[3-8\log[1-4x_\ell^2]\left(1+\log[x_\ell^2]\right)
-8{\rm Li}_2[1-4x_\ell^2]-16x_\ell^2(2+3x_\ell^2)+16x_\ell^2\left(1+2x_\ell^2\right)\log4+\frac{4\pi^2}{3}\right]\no\\
&&\simeq\frac{\alpha}{4\pi}
\left[3+16x_\ell^2\left(2-\log4\right)\right]\simeq0.002+0.006x_\ell^2.
\eeqa
We further estimate the two-loop non-universality effect to be
\beqa
\delta_{2-{\rm loop}}=\mathcal{O}\left(\alpha^2x_\tau^2\right)\simeq10^{-5}.
\eeqa
We therefore consider, for all practical purposes,
\beqa
\Gamma\left[V\to\ell\ell\right]&&=4\pi\alpha^2\left(\frac{f_V^2}{m_V}\right)Q_q^2(1-4x_l^2)^{1/2}\left(1+2x_l^2\right)\left[1+\frac{3\alpha}{4\pi}+\frac{4\alpha x_\ell^2}{\pi}\left(2-\log4\right)\right],
\eeqa
and
\beqa
R^V_{\tau/\ell}&&=(1-4x_\tau^2)^{1/2}\left(1+2x_\tau^2\right)\left[1+\frac{4\alpha x_\tau^2}{\pi}\left(2-\log[4]\right)\right].
\eeqa
%

\section{Other EFT operators}\label{app:ops}

\subsection{$Z$ mediated operators}\label{App:Zpole}

Here we consider the following set of dimension six operators
\beqa
\mathcal{L}_{Hl}=&&iC_{H\ell}^1\left(H^\dagger\overleftrightarrow{D}_\mu H\right)\left(\bar L\gamma^\mu L\right)+iC_{H\ell}^3\left(H^\dagger\overleftrightarrow{D}_\mu\tau^aH\right)\left(\bar L\gamma^\mu\tau_aL\right)+iC_{He}\left(H^\dagger\overleftrightarrow{D}_\mu H\right)\left(\bar e_R\gamma^\mu e_R\right),
\eeqa
which affect $R^V_{\tau/\ell}$ by modifying the $Z\tau\tau$ vertex. Their contributions to the $\psi$ and $\Upsilon$ leptonic decays are given by
\beqa
A_V^{q\ell}&&=-4\pi\alpha Q_q-\frac{m_V^2g_V^q}{2}\left(C_{H\ell}^1+C_{H\ell}^3/4+C_{He}\right)\left(\frac{1}{1-\epsilon}\right)\simeq-4\pi\alpha Q_q+\frac{m_V^2g_V^q}{v^2}\left(\delta g_{L}^{Z\tau}+\delta g_{R}^{Z\tau}\right),\no\\
B_V^{q\ell}&&=\frac{m_V^2g_V^q}{2}\left(C_{H\ell}^1+C_{H\ell}^3/4-C_{He}\right)\left(\frac{1}{1-\epsilon}\right)\simeq-\frac{m_V^2g_V^q}{v^2}\left(\delta g_{L}^{Z\tau}-\delta g_{R}^{Z\tau}\right),
\eeqa
where, as before, $\epsilon=m_V^2/m_Z^2$ and $g_v^q=T_3^q-2Q_qs_W^2$, and for the $Z\tau\tau$ vertex corrections we follow the definitions of~\cite{Efrati:2015eaa}. The consistency of LEP data with the SM prediction requires $\delta g_{L,R}^{Z\tau}\leq10^{-3}$ at $2\sigma$, which, in turn, can modify $R^V_{\tau/\ell}$ by, at most, $10^{-5}-10^{-6}$. This effect is negligible given the current and future experimental sensitivity.

\subsection{Dipole operator}\label{App:Dpole}

Here we consider the dimension six dipole operator
\beqa
\mathcal{L}_D=&&\sqrt{4\pi\alpha}C_D^{\ell}\bar L\sigma^{\mu\nu}e_R HF_{\mu\nu}+{\rm h.c.}.
\eeqa
Its contribution to the $\psi$ and $\Upsilon$ leptonic decays is given by
\beqa
A_V^{q\ell}&&=-4\pi\alpha Q_q+16\pi\alpha Q_q\frac{vm_\ell}{\sqrt{2}}{\rm Re}\left[C_D^{\ell}\right]=-4\pi\alpha Q_q-4\pi\alpha Q_q\Delta a_\ell,\no\\
C_V^{q\ell}&&=8\pi\alpha Q_q\frac{vm_V}{\sqrt{2}}{\rm Re}\left[C_D^{\ell}\right]=-2\pi\alpha Q_q\Delta a_\ell\left(\frac{m_V}{m_\ell}\right),\no\\
D_V^{\ell}&&=8\pi\alpha Q_q\frac{vm_V}{\sqrt{2}}{\rm Im}\left[C_D^{\ell}\right]=4\pi\alpha Q_qm_V\left(\frac{d_\ell}{e}\right),
\eeqa
where $\Delta a_\ell$ and $d_\ell$ are the leptonic magnetic and electric moments, respectively. The $\tau$ constraints read
\beqa
-0.052&&\leq\Delta a_\tau\leq0.013,\no\\
-0.0011\,{\rm GeV}^{-1}&&\leq\left(\frac{d_\tau}{e}\right)\leq0.0023\,{\rm GeV}^{-1}.
\eeqa
These bounds are, however, not strong enough to suppress the dipole contribution to $R_{\tau/\ell}^V$.


\end{document}